\def\arxivversion{1}
\ifdefined\arxivversion
  \documentclass[sigconf,nonacm]{acmart}
  \setkeys{acmart.cls}{balance=false}
  \setcopyright{none}
  \makeatletter
  \patchcmd{\@mkauthors@iii}{\par}{\relax}
    {}{\ClassError{acmart}{Cannot apply compact author layout}{Check the acmart author renderer.}}
  \patchcmd{\@mkauthors@iii}{\par}{\quad}
    {}{\ClassError{acmart}{Cannot apply compact author layout}{Check the acmart author renderer.}}
  \patchcmd{\@mkauthors@iii}
    {\par\bgroup}
    {\quad\bgroup}
    {}{\ClassError{acmart}{Cannot apply compact email layout}{Check the acmart author renderer.}}
  \patchcmd{\@typeset@author@bx}
    {\@authorfont\@currentauthors\par}
    {\parbox{0.75\linewidth}{\centering\@authorfont\@currentauthors\par}\par}
    {}{\ClassError{acmart}{Cannot set author group width}{Check the acmart author renderer.}}
  \makeatother
\else
  \documentclass[manuscript,review,anonymous]{acmart}
\fi
\usepackage{multirow} 
\usepackage{array,longtable} 
\usepackage{placeins} 

\graphicspath{{figures/}{./}}
\AtBeginDocument{%
  }

\ifdefined\arxivversion
\else
\setcopyright{acmlicensed}
\copyrightyear{2027}
\acmYear{2027}
\acmDOI{XXXXXXX.XXXXXXX}
\acmConference[CHI '27]{CHI Conference on Human Factors in Computing Systems}{May 10--14,
  2027}{Pittsburgh, PA, USA}
\acmISBN{978-1-4503-XXXX-X/2027/05}
\fi

\begin{document}

\ifdefined\arxivversion
  \title[NarrativeSteward: Delegation, Guidance, and Verification]{NarrativeSteward: Coordinating Delegation, Guidance, and Verification in Agent-Assisted Interactive Narrative Authoring}
\else
  \title[NarrativeSteward: Delegation, Guidance, and Verification]{NarrativeSteward: Coordinating Delegation, Guidance, and Verification in AI-Assisted Interactive Narrative Authoring}
\fi
\newcommand{\StudyOwnershipEqualN}{3}
\newcommand{\StudyOwnershipNegativeN}{3}
\newcommand{\StudyUnderstandingEqualN}{2}
\newcommand{\StudyPreferenceInsufficientN}{3}
\newcommand{\StudyCompleteN}{12}
\newcommand{\StudyAgentFirstN}{8}
\newcommand{\StudyGFlowFirstN}{4}
\newcommand{\StudyExternalFailureN}{1}
\newcommand{\StudyAgentTurnParticipants}{12}
\newcommand{\StudyAgentTurnEvents}{179}
\newcommand{\StudyDirectEditParticipants}{3}
\newcommand{\StudyDirectEditEvents}{6}
\newcommand{\StudyReviewParticipants}{7}
\newcommand{\StudyReviewEvents}{46}
\newcommand{\StudyRevertParticipants}{1}
\newcommand{\StudyRevertEvents}{1}
\newcommand{\StudyOwnershipGFlowMedian}{5.5}
\newcommand{\StudyOwnershipGFlowMin}{2}
\newcommand{\StudyOwnershipGFlowMax}{7}
\newcommand{\StudyOwnershipAgentMedian}{5}
\newcommand{\StudyOwnershipAgentMin}{1}
\newcommand{\StudyOwnershipAgentMax}{6}
\newcommand{\StudyOwnershipDiffMedian}{+0.5}
\newcommand{\StudyOwnershipDiffMin}{-3}
\newcommand{\StudyOwnershipDiffMax}{+3}
\newcommand{\StudyOwnershipPositiveN}{6}
\newcommand{\StudyUnderstandingPositiveN}{9}
\newcommand{\StudyUnderstandingNegativeN}{1}
\newcommand{\StudyFrustrationLowerN}{6}
\newcommand{\StudyFrustrationHigherN}{3}
\newcommand{\StudyPreferGFlowN}{7}
\newcommand{\StudyPreferAgentN}{1}
\newcommand{\StudyPolishGFlowN}{8}
\newcommand{\StudyDeliveryGFlowN}{11}
\newcommand{\StudySUSMedian}{73.8}
\newcommand{\StudySUSMin}{60}
\newcommand{\StudySUSMax}{95}
\newcommand{\StudyValidationParticipants}{11}
\newcommand{\StudyValidationEvents}{34}
\newcommand{\StudyPlaytestParticipants}{12}
\newcommand{\StudyFormulateNS}{6}
\newcommand{\StudyFormulateAgent}{3.5}
\newcommand{\StudyFormulateN}{12}
\newcommand{\StudyFormulatePositive}{11}
\newcommand{\StudyFormulateEqual}{1}
\newcommand{\StudyFormulateNegative}{0}
\newcommand{\StudyFormulateEffect}{1}
\newcommand{\StudyFormulateP}{0.003}
\newcommand{\StudyExpressNS}{6}
\newcommand{\StudyExpressAgent}{4}
\newcommand{\StudyExpressN}{12}
\newcommand{\StudyExpressPositive}{9}
\newcommand{\StudyExpressEqual}{3}
\newcommand{\StudyExpressNegative}{0}
\newcommand{\StudyExpressEffect}{1}
\newcommand{\StudyExpressP}{0.008}
\newcommand{\StudyInspectChangesNS}{6.5}
\newcommand{\StudyInspectChangesAgent}{3}
\newcommand{\StudyInspectChangesN}{12}
\newcommand{\StudyInspectChangesPositive}{12}
\newcommand{\StudyInspectChangesEqual}{0}
\newcommand{\StudyInspectChangesNegative}{0}
\newcommand{\StudyInspectChangesEffect}{1}
\newcommand{\StudyInspectChangesP}{0.002}
\newcommand{\StudyRequestFitNS}{5}
\newcommand{\StudyRequestFitAgent}{4.5}
\newcommand{\StudyRequestFitN}{12}
\newcommand{\StudyRequestFitPositive}{4}
\newcommand{\StudyRequestFitEqual}{6}
\newcommand{\StudyRequestFitNegative}{2}
\newcommand{\StudyRequestFitEffect}{0.52}
\newcommand{\StudyRequestFitP}{0.375}
\newcommand{\StudyUnderstandingNS}{6}
\newcommand{\StudyUnderstandingAgent}{4}
\newcommand{\StudyUnderstandingN}{12}
\newcommand{\StudyUnderstandingPositive}{9}
\newcommand{\StudyUnderstandingEqual}{2}
\newcommand{\StudyUnderstandingNegative}{1}
\newcommand{\StudyUnderstandingEffect}{0.91}
\newcommand{\StudyUnderstandingP}{0.029}
\newcommand{\StudyStabilityNS}{6}
\newcommand{\StudyStabilityAgent}{4.5}
\newcommand{\StudyStabilityN}{8}
\newcommand{\StudyStabilityPositive}{7}
\newcommand{\StudyStabilityEqual}{0}
\newcommand{\StudyStabilityNegative}{1}
\newcommand{\StudyStabilityEffect}{0.89}
\newcommand{\StudyStabilityP}{0.062}
\newcommand{\StudyOverviewNS}{7}
\newcommand{\StudyOverviewAgent}{4}
\newcommand{\StudyOverviewN}{12}
\newcommand{\StudyOverviewPositive}{12}
\newcommand{\StudyOverviewEqual}{0}
\newcommand{\StudyOverviewNegative}{0}
\newcommand{\StudyOverviewEffect}{1}
\newcommand{\StudyOverviewP}{0.002}
\newcommand{\StudyLocateNS}{6}
\newcommand{\StudyLocateAgent}{4}
\newcommand{\StudyLocateN}{12}
\newcommand{\StudyLocatePositive}{12}
\newcommand{\StudyLocateEqual}{0}
\newcommand{\StudyLocateNegative}{0}
\newcommand{\StudyLocateEffect}{1}
\newcommand{\StudyLocateP}{0.002}
\newcommand{\StudyRefinementNS}{5}
\newcommand{\StudyRefinementAgent}{4}
\newcommand{\StudyRefinementN}{12}
\newcommand{\StudyRefinementPositive}{7}
\newcommand{\StudyRefinementEqual}{4}
\newcommand{\StudyRefinementNegative}{1}
\newcommand{\StudyRefinementEffect}{0.72}
\newcommand{\StudyRefinementP}{0.094}
\newcommand{\StudyConfidenceNS}{6}
\newcommand{\StudyConfidenceAgent}{5}
\newcommand{\StudyConfidenceN}{12}
\newcommand{\StudyConfidencePositive}{8}
\newcommand{\StudyConfidenceEqual}{3}
\newcommand{\StudyConfidenceNegative}{1}
\newcommand{\StudyConfidenceEffect}{0.91}
\newcommand{\StudyConfidenceP}{0.031}
\newcommand{\StudyRecoveryNS}{6}
\newcommand{\StudyRecoveryAgent}{4.5}
\newcommand{\StudyRecoveryN}{8}
\newcommand{\StudyRecoveryPositive}{6}
\newcommand{\StudyRecoveryEqual}{0}
\newcommand{\StudyRecoveryNegative}{2}
\newcommand{\StudyRecoveryEffect}{0.53}
\newcommand{\StudyRecoveryP}{0.211}
\newcommand{\StudyKeepNS}{6}
\newcommand{\StudyKeepAgent}{3.5}
\newcommand{\StudyKeepN}{10}
\newcommand{\StudyKeepPositive}{10}
\newcommand{\StudyKeepEqual}{0}
\newcommand{\StudyKeepNegative}{0}
\newcommand{\StudyKeepEffect}{1}
\newcommand{\StudyKeepP}{0.008}
\newcommand{\StudyCheckNS}{5}
\newcommand{\StudyCheckAgent}{3}
\newcommand{\StudyCheckN}{9}
\newcommand{\StudyCheckPositive}{9}
\newcommand{\StudyCheckEqual}{0}
\newcommand{\StudyCheckNegative}{0}
\newcommand{\StudyCheckEffect}{1}
\newcommand{\StudyCheckP}{0.012}
\newcommand{\StudyNoPreferenceN}{1}
\newcommand{\StudyMainSignificantN}{9}
\newcommand{\StudyGateExcludedN}{0}
\newcommand{\StudyGraphsMedian}{6.5}
\newcommand{\StudyGraphsRatedN}{12}
\newcommand{\StudyGraphsUsedN}{12}
\newcommand{\StudyLocateFeatureMedian}{7}
\newcommand{\StudyLocateFeatureRatedN}{7}
\newcommand{\StudyLocateFeatureUsedN}{7}
\newcommand{\StudyContentMedian}{7}
\newcommand{\StudyContentRatedN}{11}
\newcommand{\StudyContentUsedN}{11}
\newcommand{\StudyDirectMedian}{6}
\newcommand{\StudyDirectRatedN}{4}
\newcommand{\StudyDirectUsedN}{4}
\newcommand{\StudyUndoMedian}{4}
\newcommand{\StudyUndoRatedN}{3}
\newcommand{\StudyUndoUsedN}{3}
\newcommand{\StudyVerificationHelpMedian}{7}
\newcommand{\StudyVerificationHelpRatedN}{11}
\newcommand{\StudyVerificationHelpUsedN}{11}
\newcommand{\StudyPlaytestHelpMedian}{7}
\newcommand{\StudyPlaytestHelpRatedN}{12}
\newcommand{\StudyPlaytestHelpUsedN}{12}
\newcommand{\StudyRequestedNS}{12}
\newcommand{\StudyConsideredOnlyNS}{0}
\newcommand{\StudyUnfinishedNS}{2}
\newcommand{\StudyExpectedReviewNS}{0}
\newcommand{\StudyExpectedCommunicationNS}{0}
\newcommand{\StudyTimeLimitedNS}{1}
\newcommand{\StudyNoFurtherInvestmentNS}{1}
\newcommand{\StudyRequestedAgent}{12}
\newcommand{\StudyConsideredOnlyAgent}{0}
\newcommand{\StudyUnfinishedAgent}{5}
\newcommand{\StudyExpectedReviewAgent}{4}
\newcommand{\StudyExpectedCommunicationAgent}{2}
\newcommand{\StudyTimeLimitedAgent}{0}
\newcommand{\StudyNoFurtherInvestmentAgent}{2}

\author{Wenjin~Wang}
\email{vinjinwang@tencent.com}
\author{Jiazhen~Lei}
\author{Yuxin~Sha}
\author{Nuwa~Xi}
\author{Meng~Zhao}
\author{Xingxi~Yin}
\author{Qi~Liu}
\author{Yuliang~Shen}
\author{Zixun~Sun}
\authornote{Corresponding author.}
\email{zixunsun@tencent.com}

\affiliation{%
  \institution{Interactive Entertainment Group, Tencent Inc}
  \city{Shanghai}
  \country{China}
}

\renewcommand{\shortauthors}{Wang et al.}

\begin{abstract}
Autonomous AI agents can turn authors’ goals into interactive narratives by independently organizing and carrying out generation and revision. As agents generate and revise extensive content, authors struggle to grasp its overall structure, local details, and relationships, complicating continued guidance. We present NarrativeSteward, an authoring environment that organizes outlines, worldbuilding, and narrative graphs as linked artifacts for agent implementation and author guidance. Agent dialogue and project-wide structural review help authors understand the evolving work and guide local and cross-layer revisions, while change records and execution verification help authors assess the resulting work. Technical tests validated the system’s change records, recovery mechanisms, and execution diagnostics. In a 12-participant within-subject study, NarrativeSteward supported easier formulation of revision requests and inspection of changes, and greater perceived understanding of changes and story structure, than general-purpose agents. Qualitative findings show how reviewing the work and feedback helps authors develop requirements and guide subsequent delegation.
\ifdefined\arxivversion
We open-source NarrativeSteward at \url{https://github.com/Tencent/NarrativeSteward}.
\fi
\end{abstract}


\begin{teaserfigure}
  \centering
  \ifdefined\arxivversion
  \includegraphics[width=0.72\textwidth]{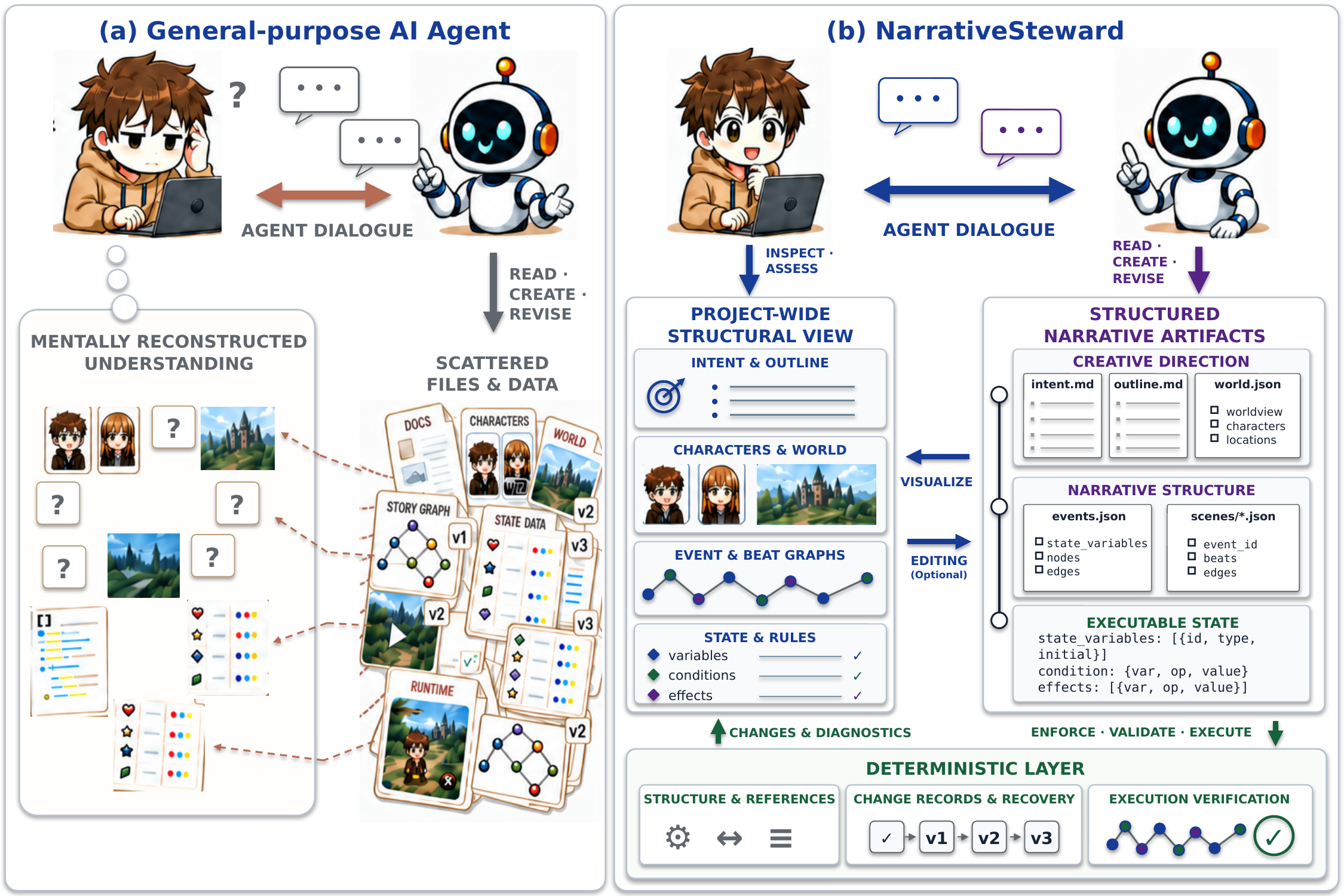}
  \else
  \includegraphics[width=\textwidth]{figure1_revision_20260910/figure1_export_compatible.pdf}
  \fi
  \caption{Interactive narrative authoring with general-purpose AI agents and NarrativeSteward.
  (a) With general-purpose AI agents, authors can delegate generation and revision but face the challenge of understanding an evolving narrative across scattered files and representations.
  (b) NarrativeSteward coordinates delegation, guidance, and verification through linked narrative artifacts.
  The agent organizes generation and revision across these artifacts, while a project-wide structural view helps authors understand their content and relationships.
  Together with change records and execution diagnostics, this view helps authors assess the evolving work and guide subsequent revisions through dialogue.}
  \Description{A two-panel comparison of interactive narrative authoring.
  The left panel shows an author and a general-purpose AI agent in dialogue, with documents, graphs, state data, and runtime outputs feeding an incomplete mental representation of the story.
  The right panel shows NarrativeSteward: an author and agent exchange requests and responses above a project-wide structural view and linked narrative artifacts.
  A downward arrow from the author to the structural view indicates inspection and assessment.
  File cards and field snippets depict the structured project data on the right; a leftward visualization arrow connects them to outline, character, graph, and state-rule representations in the structural view.
  The agent reads, creates, and revises artifacts; the structural view displays their content and relationships and provides optional direct editing.
  A deterministic layer maintains structural references, change records and recovery, and execution verification, with changes and diagnostics returning to the structural view.}
  \label{fig:system_overview}
\end{teaserfigure}


\maketitle

\section{Introduction}
\label{sec:intro}

Recent advances in agent planning and sustained task execution, exemplified by Codex and Claude Code, are expanding the scope of work that users can delegate~\cite{openai2026gpt53codex,anthropic2026claudeopus46}.
These agents can inspect project files, select tools, and adjust their actions in response to execution feedback.
For interactive narrative authors, these capabilities make it possible to delegate substantial implementation, from creating a playable first draft to revising events, choices, and state rules.
As extensive content is generated, authors face the challenge of grasping the overall structure, local details, and their relationships.
To guide further creation, they need to determine what the work currently contains, how it reflects their intentions, and which aspects warrant another request.
Understanding the developing work is therefore a continuing part of delegation.

Interactive narrative connects these judgments across several levels of creative decisions.
Outlines and worldbuilding give direction and context to events, while local choices and state changes determine what players can encounter along different paths~\cite{riedl2013interactive,hargood2022authoring}.
An author who wants early decisions to matter may need to examine the branching structure, the content of individual choices, and their later consequences before deciding how to revise the story.
These concerns involve both narrative meaning and execution: a branch can be reachable yet feel inconsequential, while a desired continuation can remain inaccessible under the authored rules~\cite{otto2023dendryscope,veloso2021validating}.
Supporting author guidance requires connecting an understanding of the whole work with inspection of its constituent parts.

Interactive narrative authoring tools have long helped authors organize and edit branching content and its connections~\cite{friedhoff2013untangling,hargood2022authoring}.
More recent systems integrate LLMs into graph editing, content generation, and narrative analysis~\cite{calderwood2022spinning,mishra2025whatif,li2026exploring}.
Other approaches connect outlines or author-defined specifications with generated narrative experiences, including visual support for inspecting possible storylines~\cite{lu2025whatelse,wu2026orchid,wang2026elsewise}.
Generation frameworks also translate narrative requirements into branching structures, scenes, and playable content~\cite{kumaran2023scenecraft,kumaran2024narrativegenie,leandro2024geneva}, with some incorporating automated checking and repair~\cite{rogosch2026automated,puchalski2026symmetry}.
These approaches expand creative support through authoring interfaces, editing and analysis functions, and structured generation procedures.

Delegation to autonomous agents introduces a different division of work: authors can entrust the organization and execution of implementation to an agent while continuing to determine the story's direction.
Requests can range from a local refinement to a broader change in how the narrative unfolds, with the agent deciding which artifacts to inspect and modify and how to proceed from feedback.
How to support continuing author guidance under this form of delegation remains underexplored in interactive narrative authoring.

\begin{samepage}
\noindent Our central design question is: \textbf{how can authors delegate interactive narrative implementation while understanding the evolving work and continuing to guide its refinement?}
\par
\end{samepage}

To address this question, we present \textbf{NarrativeSteward}, an authoring environment that supports autonomous implementation and continuing author guidance (Figure~\ref{fig:system_overview}).
Its design centers on \textbf{linked narrative artifacts}, including outlines, worldbuilding, and narrative graphs, which provide a shared basis for agent implementation, author review, and feedback.
The agent can organize local and cross-layer changes around these artifacts without a fixed authoring sequence.

\begin{figure*}[!t]
  \centering
  \includegraphics[width=\linewidth]{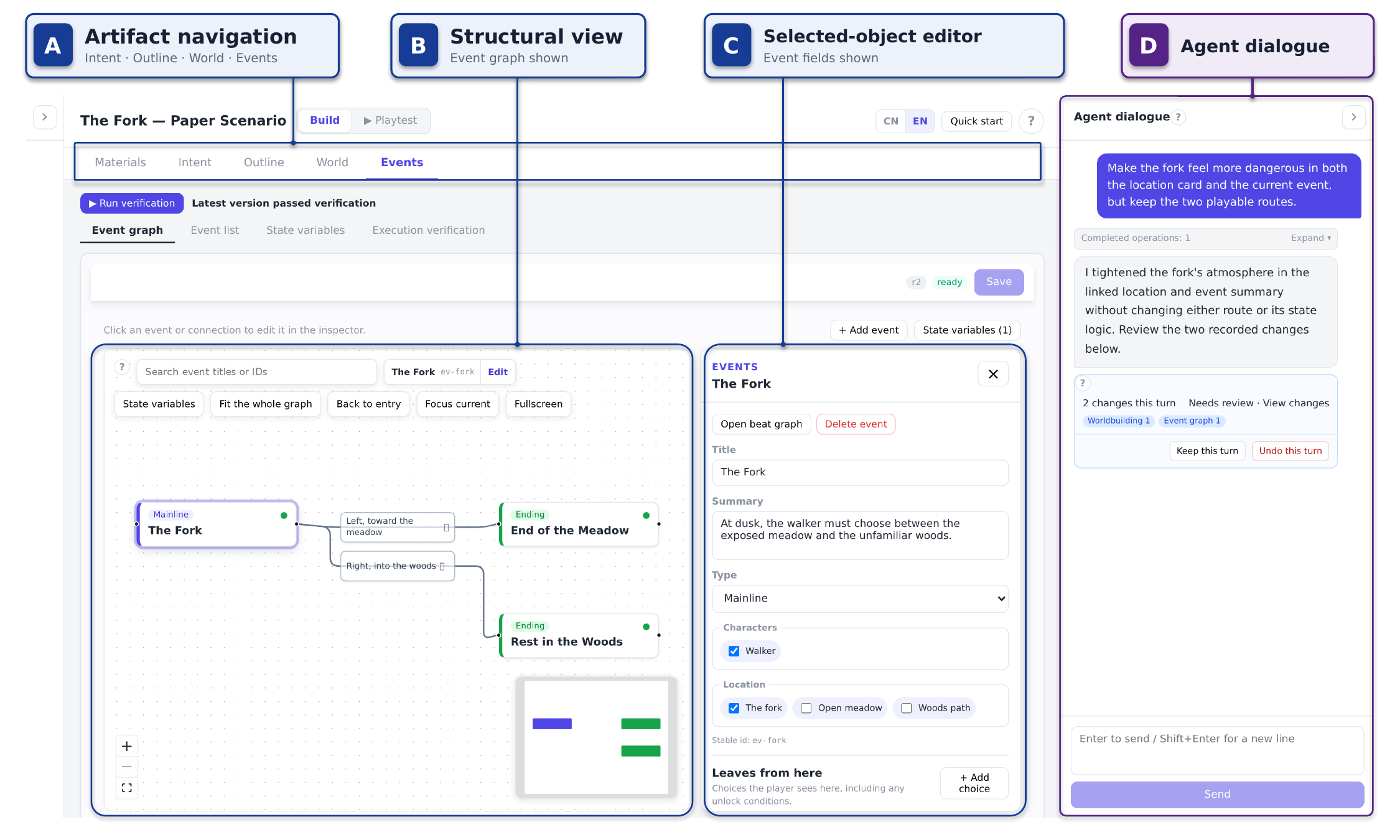}
  \caption{\textbf{The NarrativeSteward authoring interface.}
  Structural review, local inspection, and agent dialogue are connected in a shared workspace.
  Artifact navigation (A) provides access to the narrative layers.
  The structural view (B) shows the branching relationships of the selected event, \emph{The Fork}, while the selected-object editor (C) exposes its content and references.
  In the agent dialogue (D), the author requests a more dangerous atmosphere in both the location card and the event while retaining two playable routes.
  A change record summarizes modifications to worldbuilding and the event graph, connecting the request with the resulting changes.}
  \Description{An annotated screenshot of NarrativeSteward with four labeled regions.
  Artifact navigation (A) appears above the structural view (B) and selected-object editor (C), with agent dialogue (D) on the right.
  The event graph shows The Fork branching to End of the Meadow and Rest in the Woods.
  The editor displays The Fork's summary, character, and location references.
  The dialogue requests a more dangerous atmosphere in the location card and event while preserving two playable routes.
  Below the response, a change record lists one worldbuilding change and one event-graph change, with View changes, Keep this turn, and Undo this turn controls.}
  \label{fig:system_anatomy}
\end{figure*}

NarrativeSteward combines \textbf{agent dialogue} with \textbf{project-wide structural review} to help authors understand the evolving work and guide further revisions (Figure~\ref{fig:system_anatomy}).
Authors can examine content and relationships in the structural view, bring concerns into dialogue, and develop subsequent requests.
Change records identify actual modifications and connect them to the affected artifacts, while execution verification identifies unreachable content and dead ends.
Together, these interactions let authors assess the work in context and decide what the agent should do next.

We used controlled technical tests to validate the system's change records, recovery mechanisms, and execution diagnostics.
A within-subject mixed-method study with \StudyCompleteN{} participants compared NarrativeSteward with general-purpose agents, examining delegation experience, perceived understanding, and how authors used feedback to judge results and guide revisions.
Both conditions allowed authors to delegate implementation; the comparison examined authoring experience in the two environments as a whole.
The user study addresses three research questions:

\begin{description}
  \item[RQ1:] How does NarrativeSteward shape authors' experience and effort when delegating interactive narrative implementation and revision to an agent?
  \item[RQ2:] How does the combination of dialogue and structured artifact views shape authors' experience of understanding and continuing to refine their work?
  \item[RQ3:] How does system feedback inform authors' judgments of implementation results and their subsequent revision decisions?
\end{description}

Compared with general-purpose agents, NarrativeSteward supported easier formulation of revision requests and inspection of changes, alongside greater perceived understanding of changes and story structure.
Qualitative findings show how structural review helped authors identify concerns and develop them through dialogue into subsequent requests.
Authors also used execution diagnostics and agent feedback on narrative consistency to decide what to inspect and revise next.
These accounts show how guidance develops through examining the work as well as specifying initial goals.
Differences in review preferences and appreciation of general-purpose agents' flexibility motivate support for choosing when and how closely to inspect the work and how to continue its development.

This work makes three contributions:

\begin{enumerate}
  \item \textbf{A design and implemented system for coordinating delegation, guidance, and verification through linked narrative artifacts.}
  NarrativeSteward connects autonomous implementation, agent dialogue, structural review, and feedback around the same evolving work, supporting local and cross-layer author guidance.
  \item \textbf{Technical and user-study evidence for this authoring approach.}
  Controlled tests evaluate change records, recovery mechanisms, and execution diagnostics; the comparative study examines delegation experience, perceived understanding, and revision judgments in interactive narrative creation.
  \item \textbf{Insights into how authors sustain guidance during delegated creation.}
  The findings explain how reviewing narrative artifacts and feedback helps authors develop subsequent requests, and motivate adaptable review and flexible support for continuing creation.
\end{enumerate}

\section{Related Work}
\label{sec:related}

\subsection{Interactive Narrative Authoring}
\label{sec:rw-authoring}

Interactive narrative authoring involves designing choices and consequences across multiple possible experiences~\cite{murray1997hamlet,koenitz2023understanding,riedl2013interactive,mawhorter2014choice}.
As stories grow, authors need representations that help them organize content and reason about connections beyond an individual passage~\cite{hargood2022authoring}.
Branching editors such as Twine, storylet-based structures, and progression visualizations offer different ways to organize and inspect narrative possibilities~\cite{friedhoff2013untangling,kreminski2018storylets,kreminski2020why,partlan2019evaluation,carstensdottir2020progression}.
These approaches make the structure of an interactive work available for author decisions alongside its prose.

LLM-assisted tools connect this structural work with generation and analysis.
Spindle lets authors alternate between writing Twine passages themselves and requesting individual or batch generation from a language model~\cite{calderwood2022spinning}.
WhatIF combines a branching editor with generated suggestions and narrative critiques~\cite{mishra2025whatif}.
CoNoder provides text-based discussion of the narrative graph and a Function Mode that directly adds or revises graph elements referenced by the author; ripple-effect analysis and simulated-reader feedback help authors consider the consequences of edits~\cite{li2026exploring}.
These systems support an iterative authoring process in which authors select content, request assistance, and use the resulting material or analysis to continue editing.

Generation frameworks allow authors to express requirements at a higher level while the system produces detailed narrative content.
GENEVA turns a story description and constraints into branching storylines and a graph for inspecting their structure~\cite{leandro2024geneva}.
SceneCraft translates scene goals, characters, and locations into branching dialogue and game-ready scenes, while NarrativeGenie organizes narrative beats, event dependencies, and game scripts into playable content~\cite{kumaran2023scenecraft,kumaran2024narrativegenie}.
Here, generation is organized through defined transformations from author input to narrative structure and implementation.

Another line of work helps authors shape experiences that unfold through generation.
WhatELSE connects abstract outlines with concrete story instances and variants, enabling authors to refine a narrative possibility space through both specifications and examples~\cite{lu2025whatelse}.
Orchid-Creator uses author-constructed cards, a story graph, and rules to guide runtime narration~\cite{wu2025orchid,wu2026orchid}, while DiaryPlay helps authors turn linear accounts into interactive vignettes~\cite{xu2026diaryplay}.
Elsewise visualizes sampled playthroughs along author-selected narrative dimensions, helping authors examine how a story can unfold and revise its world, characters, and rules~\cite{wang2026elsewise}.
In these settings, authors shape the specifications and possibilities from which narrative experiences emerge.

Our research concerns author guidance when the organization of implementation is itself delegated.
An autonomous agent can determine which parts of a narrative project to inspect and change, and continue acting on the results, while authors introduce further goals and refinements.
NarrativeSteward organizes linked artifacts as the common basis for this implementation and for authors' understanding and subsequent direction.

\subsection{Autonomous Delegation and Structured Work Environments}
\label{sec:rw-agents}
\label{sec:rw-interaction}

Tool-using agents enable users to delegate work that involves acting in an environment and responding to its changing state~\cite{wang2024survey}.
ReAct connects reasoning with actions and observations, while software agents such as SWE-agent and OpenHands navigate repositories, modify files, and test changes~\cite{yao2023react,yang2024sweagent,wang2025openhands}.
These mechanisms support a division of work in which users specify outcomes and the agent develops and carries out implementation steps.

Creative systems also organize substantial implementation through plans and structured outputs~\cite{mirowski2023cowriting}.
In game prototyping, DreamGarden combines a hierarchical task plan with implementation modules and compilation or execution feedback~\cite{earle2025dreamgarden}.
RPGAgent coordinates narrative, spatial, and gameplay generation through specialized agents and structured intermediate data~\cite{zhang2026rpgagent}.
These examples illustrate how the organization of a development process can support coordinated creative production.

Greater automation also changes what users need to understand and review.
A study comparing Copilot with OpenHands found that participants understood the inline assistant's output better even though the agent condition improved task correctness and reduced hands-on time~\cite{chen2026code}.
Research on AI-assisted programming likewise documents effort spent reading and validating generated suggestions, and shifts between using assistance for acceleration and exploration~\cite{mozannar2024reading,barke2023grounded}.
These findings motivate interfaces that help users interpret the developing work while delegating its implementation.

Structured workspaces connect requests to concrete objects and expose results for review~\cite{bernstein2010soylent,kreminski2020germinate,chung2022talebrush}.
Wordcraft embeds open-ended writing requests in an editor; DirectGPT uses selection and direct manipulation to identify the objects and operations of a request~\cite{yuan2022wordcraft,masson2024direct}.
InkSync makes proposed document changes executable and supports checking newly introduced information~\cite{laban2024beyond}.
AI Chains exposes intermediate results for inspection and revision, while creative environments such as Spellburst, Luminate, and ImaginationVellum support refinement, comparison, or revisiting generated alternatives~\cite{wu2022ai,angert2023spellburst,suh2024luminate,marquardt2025imaginationvellum}.
Across these approaches, visible representations help users relate a request to the content it concerns and judge what has changed.

For interactive narrative projects, this connection must also span the relationships among worldbuilding, events, local choices, and state rules.
NarrativeSteward links change records and execution diagnoses to narrative artifacts, allowing authors to examine the affected content together with its connections in the structural view.
Authors can use this context to judge how a change or problem relates to their creative intentions, express their concerns in dialogue, and delegate further work on the same artifacts.

\subsection{Author Guidance and Creative Participation}
\label{sec:rw-agency}

Human–AI collaboration involves decisions about when to request assistance, how to correct its actions, and how much work to delegate~\cite{horvitz1999principles,amershi2019guidelines,shneiderman2022human}.
In creative work, these decisions also concern who develops ideas and directs their evolution~\cite{lee2022coauthor,margarido2025boosting}.
DuetDraw participants preferred to lead collaboration and request explanations rather than receive unsolicited intervention~\cite{oh2018ilead}.
A two-week study of screenwriters describes creators developing their own strategies for involving AI and working independently~\cite{tang2026how}.
Studies of proactive assistance similarly show that its timing and presentation affect whether it supports ongoing work or interrupts it~\cite{pu2025assistance,chen2025need,kuo2026developer}.

How a system contributes can affect authors' sense of leading the work.
Research on proactive story writing distinguishes inserting continuations from offering ideas for exploration, with different implications for agency and ownership~\cite{yin2026proactive}.
Comparisons of model-led rewriting and human-led questioning also show that the organization of interaction shapes idea development and perceived ownership~\cite{maier2026partnering}.
Work on creative ownership and interface metaphors further examines how people understand their roles in AI-assisted production~\cite{draxler2024ghostwriter,carrera2026where,polimetla2026paradigm}.
Plotania, for example, made contribution sources clearer without a significant change in perceived creative agency~\cite{hu2026plotania}.
Together, these studies connect creative participation to the decisions authors can make, beyond awareness of who produced particular content.

Delegated implementation makes this relationship especially relevant: authors may continue to develop and assess creative decisions while the agent performs the resulting edits.
We examine how reviewing narrative artifacts and feedback supports that participation, allowing authors to identify concerns, judge proposals, and develop subsequent requests.
This focuses guidance on the ongoing direction of the work, including when authors choose to inspect it and how they express what should happen next.

\subsection{Execution Analysis and Revision Feedback}
\label{sec:rw-verification}

Interactive narrative representations connect authorial decisions to the content that players can encounter.
Drama management, narrative planning, and fragment assembly provide different ways to organize this relationship through authored beats, action models, or conditions for selecting story material~\cite{mateas2005structuring,riedl2010narrative,porteous2010applying,garbe2019storyassembler,cardonarivera2020gfi}.
Work translating prose into planning domains further connects language-based specifications with executable representations~\cite{kelly2023there}.
These approaches make narrative progression something that can be computed and examined.

Analysis tools help authors investigate the consequences of those representations.
Story Validator analyzes paths in Twine graphs and reports properties such as ending coverage and dead ends~\cite{veloso2021validating}.
DendryScope translates quality-based narratives into answer-set programs, supporting bounded playtrace queries and visual exploration~\cite{otto2023dendryscope}.
State-space visualization and conflict analysis provide additional ways to examine softlocks or problematic authored behavior~\cite{robertson2025state,kapadia2015computer}.
These tools connect computational analysis with questions that arise when authors inspect possible play.

Recent generation systems use analysis results to drive repairs.
SINE generates Ink interactive fiction from structured educational requirements, checks the resulting scripts, and passes failures to a repair agent~\cite{rogosch2026automated}.
Puchalski and Woźna-Szcześniak combine graph analysis with LLM-generated patches to repair Twine/Twee stories, encoding story state in passage names and addressing detected faults in a defined order~\cite{puchalski2026symmetry}.
Such approaches use external diagnostics to guide model revision within a generation and repair procedure.

NarrativeSteward brings execution feedback into the author's continuing guidance of delegated work.
Verification follows choices and state updates across event and beat graphs, and connects diagnoses to the narrative objects being revised.
Authors can inspect the affected content and use the diagnosis to decide what to ask the agent to change, while playtesting provides experience of selected routes.
This connection supports revision decisions that involve both whether the narrative can proceed and how the author wants it to unfold.

\section{The NarrativeSteward System}
\label{sec:system}

NarrativeSteward is a web-based authoring environment in which authors delegate the creation and revision of interactive narratives to an autonomous agent while continuing to guide the developing work.
The agent organizes and carries out implementation across linked narrative artifacts; authors use dialogue, structural views, and feedback to understand the work, assess it against their intentions, and direct further development.
The resulting narratives support branching choices, delayed consequences, and multiple endings.

\subsection{Design Objectives}
\label{sec:objectives}

When an agent determines how to carry out a creative request, authors need to understand the resulting work and decide what to ask for next.
In interactive narratives, these judgments can involve both local content and its consequences across different branches.
Building on research on narrative authoring, interaction with creative artifacts, and human–AI collaboration~\cite{li2026exploring,angert2023spellburst,masson2024direct,horvitz1999principles,amershi2019guidelines}, we translate these needs into three design objectives.

\textbf{DG1: Support delegation and review of local and cross-layer changes.}
The system should support local and cross-layer requests, allowing the agent to determine the implementation steps and affected artifacts.
It should make the actual scope and content of changes inspectable in relation to the author’s request.

\textbf{DG2: Connect focused dialogue with project-wide structural review.}
The system should support movement between focused dialogue, project-wide relationships, and local content, helping authors examine a concern at the relevant narrative levels and bring what they find back into discussion.

\begin{figure*}[!t]
  \centering
  \includegraphics[width=\linewidth]{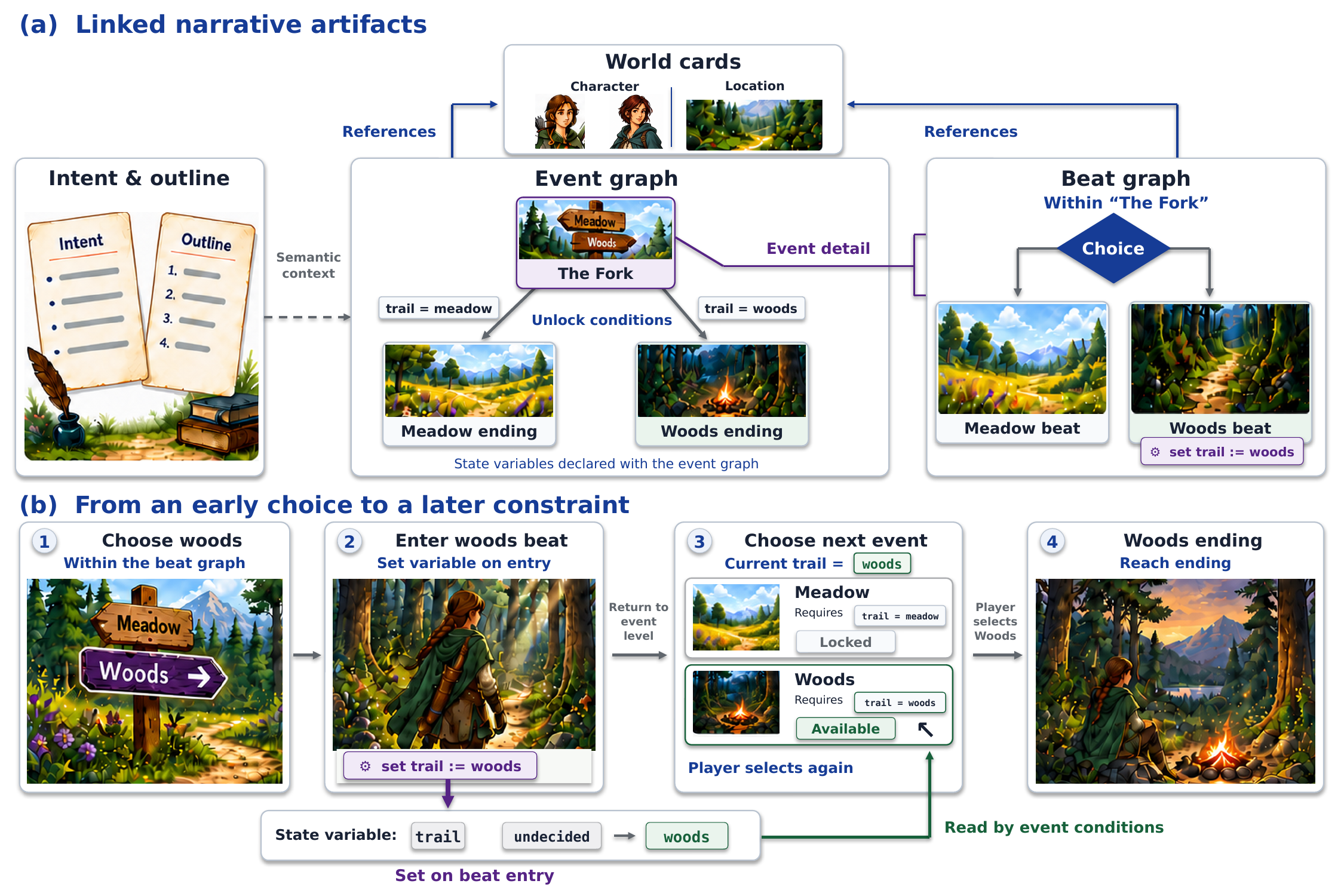}
  \caption{Linked narrative artifacts and how choices shape later events.
  (a) The event graph organizes the story’s events, while each event contains a beat graph of local narrative content and choices.
  Events and beats reference character and location cards from the worldbuilding; intent and outline provide creative context.
  These linked artifacts provide a shared basis for agent implementation and author review.
  (b) In \emph{The Fork}, the player chooses woods and enters a beat that sets the state variable \texttt{trail} to \texttt{woods}, recording that choice.
  When the event ends, conditions on outgoing event edges check this value: the woods destination becomes available, while meadow remains locked.
  The player then selects woods and reaches its ending.}
  \Description{The top panel links intent and outline to an event graph, character and location world cards, and a magnified beat graph within The Fork.
  Its meadow and woods event edges have matching conditions on trail.
  The bottom panel shows choosing woods, entering a beat that sets trail to woods, selecting the enabled woods event while meadow is locked, and reaching the woods ending.
  Purple and green arrows connect state writes and reads to the shared variable box.}
  \label{fig:linked_artifacts}
\end{figure*}

\textbf{DG3: Make feedback useful for subsequent creative decisions.}
The system should connect change records, execution diagnostics, and playtesting feedback to relevant narrative content to help authors decide what to retain, revise, or investigate further.

Authors set creative direction and choose when to review or request further work; the agent interprets their goals, organizes implementation, and acts on the relevant artifacts; the system records changes and checks structural and execution rules.

\subsection{Linked Artifacts across Dialogue and Structural Views}
\label{sec:workspace}

NarrativeSteward organizes the project around linked narrative artifacts that remain available across successive requests (Figure~\ref{fig:linked_artifacts}(a)).
The agent reads and revises these artifacts as needed for each task, while authors inspect the same work through dialogue and structural views.

Intent records the author's goals, while the outline describes the developing narrative direction.
Worldbuilding takes the form of world cards with stable identities for characters, locations, and other world knowledge.
An \emph{event graph} organizes major narrative situations and their connections.
Each event contains a \emph{per-event beat graph}, whose nodes, or \emph{beats}, hold narration, dialogue, monologue, or a choice.
State rules connect choices with later consequences: variables are declared in the event graph, beat effects update their values, and edge conditions determine which continuations are available.

\begin{figure*}[!t]
  \centering
  \includegraphics[width=\linewidth]{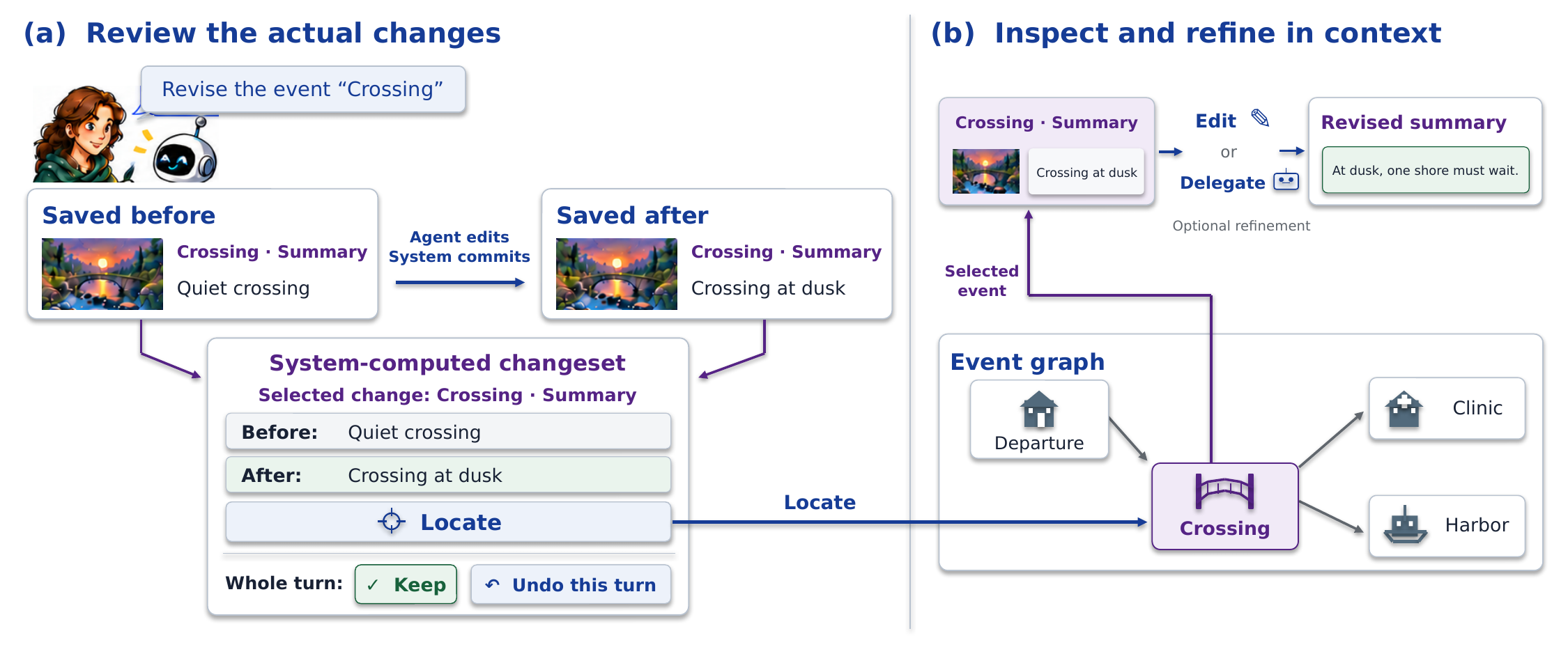}
  \caption{Reviewing saved changes and guiding further revision.
  (a) The author asks the agent to revise an event named \emph{Crossing}.
  After saving the revision, the system compares the previous and updated content to produce a \emph{changeset}, a record of the turn’s actual changes.
  The example shows a change to the event’s summary.
  \emph{Keep} retains the turn’s changes; \emph{Undo this turn} reverses them together.
  (b) \emph{Locate} selects the affected event in the graph and opens its summary field.
  The author can inspect the event and its connections, then delegate further changes or edit directly.}
  \Description{Two panels show reviewing an agent revision and continuing work on the affected event.
  Panel a compares the Crossing summary before and after a saved turn and connects both versions to a system-computed changeset.
  The selected difference has a Locate button, while Keep and Undo this turn apply to the whole turn.
  Panel b locates Crossing in an event graph and opens its Summary field.
  The author can choose Edit or Delegate as optional refinement, leading to an illustrative revised summary, At dusk, one shore must wait.}
  \label{fig:inspectable_changes}
\end{figure*}

These artifacts connect through identity references, execution dependencies, and semantic context.
Character and location references link events and beats to the corresponding world cards, connecting narrative content with its worldbuilding context.
Effects and conditions connect local actions to consequences elsewhere in the narrative.
Intent and outline provide context for interpreting these developments against the author's goals.
The system checks explicit references and execution relationships, while authors and the agent interpret themes, motivations, and other semantic connections.

The interface brings these relationships into the author's view (Figure~\ref{fig:system_anatomy}).
Artifact navigation (A) provides access to the narrative layers; the project-wide structural view (B) displays the selected layer and its connections; the selected-object editor (C) exposes local content and references; and agent dialogue (D) supports discussing the work and requesting changes.
Authors can inspect the structure and local details relevant to a discussion, then return to dialogue to express their concerns in natural language.
The agent can consult the same project artifacts to interpret these concerns and identify the relevant content and relationships.
Direct editing is also available in the object editor.

For example, Figure~\ref{fig:system_anatomy} shows a request to make a location and its associated event feel more dangerous while preserving two playable routes.
The request spans a world card and an event, and the change record indicates that both were modified.
The author can examine the event's content and branching relationships alongside the request, then decide whether the atmosphere and available routes fit the intended story.
One possible continuation is to inspect the content along both routes and consider whether the more dangerous setting is reflected in distinct risks for each route.
If those differences seem insufficient, the author can ask the agent to differentiate the risks while keeping both routes playable.

Authors can also revise an event, revisit the outline, or request changes spanning both, with the authoring sequence following their evolving creative goals.

\subsection{Executable Narrative Model}
\label{sec:representation}

Choices in an interactive narrative can affect which content becomes available later.
NarrativeSteward represents these consequences through state variables shared by the event and beat graphs.
Entering a beat applies its effects, updating variable values; conditions on outgoing edges determine which continuations are available.
When an event's beat graph finishes, control returns to the event graph, where the player selects an available destination.
Completing a terminal event concludes the story.

Figure~\ref{fig:linked_artifacts}(b) illustrates this process in \emph{The Fork}.
The variable \texttt{trail} records the walker's choice, with possible values \texttt{undecided}, \texttt{meadow}, and \texttt{woods}, initially \texttt{undecided}.
Choosing woods leads to a beat whose effect sets \texttt{trail} to \texttt{woods}; the alternative branch sets it to \texttt{meadow}.
After the beat graph finishes, the outgoing event conditions read this value.
With \texttt{trail=woods}, the woods destination is available and the meadow destination is locked.
The player then selects woods and proceeds to its ending.
This connects a choice within an event to the story's subsequent branching structure.

The event graph and each per-event beat graph are directed acyclic graphs with a single entry.
State variables are Boolean flags, finite enumerations, or bounded integers.
Beat effects assign values with \texttt{set} or increment integers with \texttt{add}, clamping the result to the declared bounds.
An edge at either level can have one condition over a single variable.
These constraints define a finite execution state space, supporting exhaustive verification.

These graphs and state rules are part of the artifacts that authors and the agent revise, with beat effects and edge conditions accessible in the corresponding object editors.
The same execution rules govern playtesting and the verification described in Section~\ref{sec:verification}.

\subsection{Delegating, Reviewing, and Guiding Revisions}
\label{sec:inspectable-edits}

Authors can delegate goals ranging from generating a playable first version to revising local content or changing several related narrative layers.
An \emph{agent turn} is the work initiated by an author request and may include multiple tool operations.
Authors can ask the agent to pause for confirmation or continue across multiple authoring tasks.

To carry out a request, the agent reads and searches the project’s narrative files, identifies relevant content and relationships, and determines which operations to perform next.
It can create or edit the outline, world cards, event graph, and per-event beat graphs, including the state rules they contain.
Retrieved content, editing results, and structural validation errors are returned to the agent, allowing it to inspect additional context, adjust its edits, or continue with another part of the task.
Thus, a single request can unfold into a sequence of actions whose details are determined by the agent in response to the developing work.
Appendix~\ref{app:implementation} describes the agent configuration and task coordination.

Successful changes are saved to the project and presented in a \emph{changeset}: a grouped record of the turn’s additions, removals, and modifications.
By comparing captured before-and-after content, the system shows what changed in the artifacts (Figure~\ref{fig:inspectable_changes}(a))~\cite{laban2024beyond}.
Authors can expand individual differences to inspect their values and assess how the revision addresses their request.
The record links this assessment to the current work through \emph{Locate}, which opens the affected object in its editor; removed objects lead to their parent or owning panel.

Locating a change allows authors to examine its surrounding content and relationships before deciding how to continue.
In Figure~\ref{fig:inspectable_changes}(b), locating the revised summary of \emph{Crossing} selects the event in the graph and opens its summary field.
The author can review the event alongside its connections, discuss the revision with the agent, and request further changes to the same work.
Direct editing provides an additional way to refine the field.

After a turn’s changes are saved, authors can select \emph{Keep} to retain them or \emph{Undo this turn} to restore the affected artifacts to their pre-turn state.
Continuing to edit or chat also marks the turn as kept.
To protect subsequent work, Undo is available only while the affected objects’ version numbers and content still match those recorded when the turn was saved.
Authors who want to retain some changes and adjust others can request a targeted revision or edit the relevant artifacts directly.

To manage each turn’s changes as a unit, the agent prepares them in an isolated draft.
After structural validation, the system saves them together and computes the changeset.
Stopping generation or a failure before saving discards the draft; if saving fails, the system restores the previous saved state.

\subsection{Execution Verification and Diagnostic Feedback}
\label{sec:verification}

\begin{figure*}[!t]
  \centering
  \includegraphics[width=\linewidth]{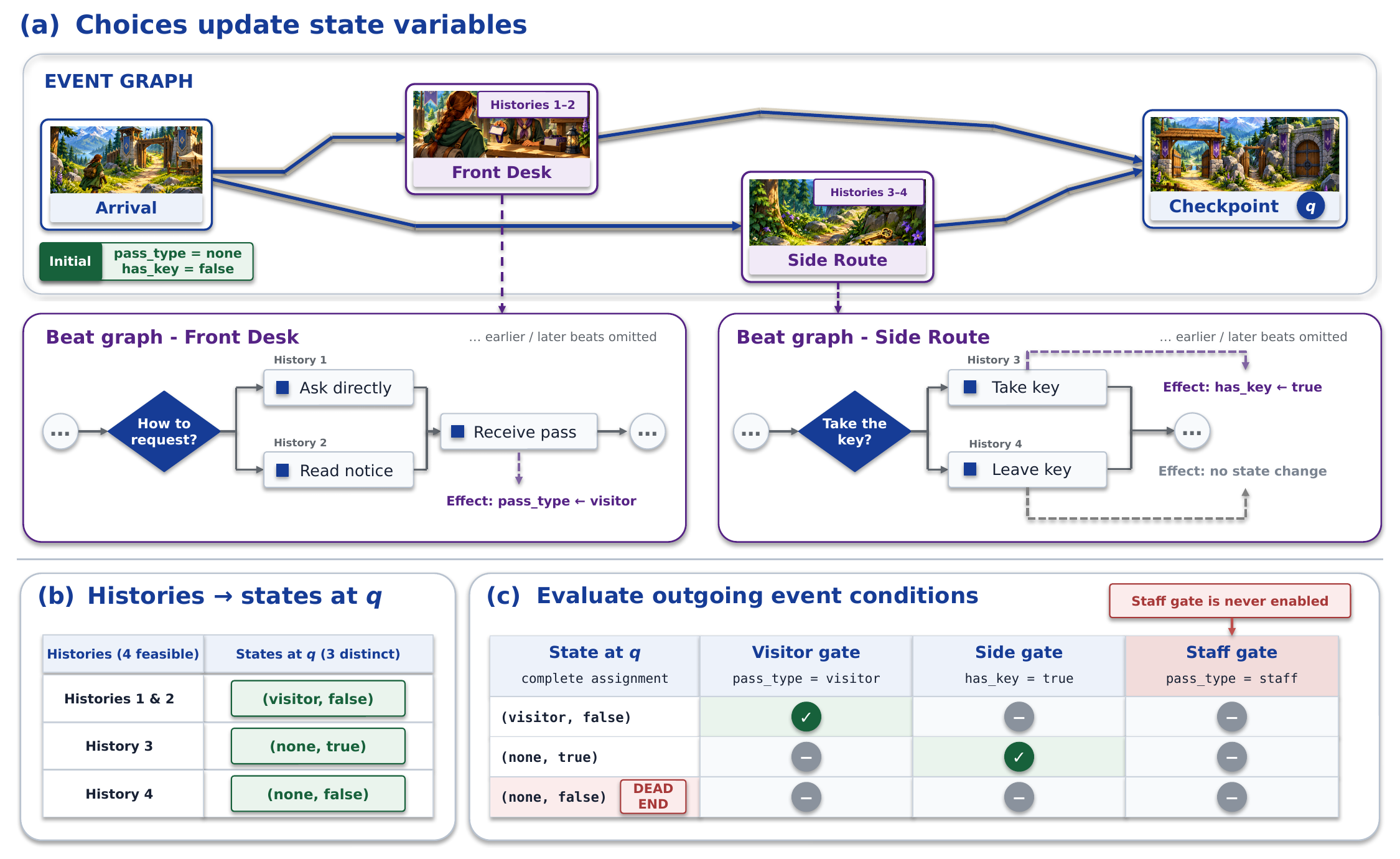}
  \caption{Execution verification across narrative choices.
  (a) Choices within the event and beat graphs determine whether the player reaches the \emph{Checkpoint} event, marked \(q\), with a visitor pass, a key, or neither.
  A history is one feasible execution from Arrival to \(q\).
  (b) Four histories produce three states: two histories yield the same visitor-pass state.
  Tuples give (\texttt{pass\_type}, \texttt{has\_key}).
  (c) Each gate column represents an outgoing edge of the checkpoint event; its condition is tested against every reachable state.
  A checkmark indicates an available exit, and a dash an unavailable one.
  With neither a pass nor a key, no exit is available, creating a dead end.
  The Staff gate is never enabled because no history supplies a staff pass.}
  \Description{Three panels show how choices in event and beat graphs produce reachable states and checkpoint diagnostics.
  Panel a shows Arrival branching through Front Desk or Side Route before reconverging at Checkpoint q.
  Purple dashed connectors expand the two events into partial beat graphs, with ellipses for omitted beats.
  At Front Desk, asking directly or reading a notice leads to receiving a visitor pass.
  On Side Route, taking the key sets has\_key to true, while leaving it changes no state.
  Panel b maps Histories 1 and 2 to (visitor, false), History 3 to (none, true), and History 4 to (none, false).
  Panel c tests each state against Visitor, Side, and Staff gate conditions.
  The visitor-pass state enables only Visitor, the key-only state enables only Side, and the state with neither enables no gate and is marked as a dead end.
  The Staff gate is never enabled.}
  \label{fig:state_propagation}
\end{figure*}

To assess whether a developing narrative remains playable, authors need to check the consequences of its choices and state rules.
NarrativeSteward provides \emph{execution verification}, an exhaustive analysis of the saved event and beat graphs under these rules~\cite{otto2023dendryscope}.
It explores reachable execution states to identify unreachable content, transitions that can never be taken, and non-ending states with no available continuation.
Authors can request this check and use its diagnostics to locate the relevant narrative objects, inspect their content and rules, and guide further revisions.

An \emph{execution state} consists of a position in the event or beat graph and the values of the shared variables.
The same position can be reached with different values, enabling different continuations.

Figure~\ref{fig:state_propagation} illustrates this process in a checkpoint story.
The variable \texttt{pass\_type} records a pass category (\texttt{none}, \texttt{visitor}, or \texttt{staff}), while \texttt{has\_key} records whether the player has a key; initially, their values are \texttt{none} and \texttt{false}.
The event graph branches from Arrival through Front Desk or Side Route before reconverging at Checkpoint, marked \(q\).
A \emph{history} is one feasible execution from Arrival to \(q\) through the event and beat graphs.
At Front Desk, asking directly or reading the notice both lead to receiving a visitor pass (Histories 1–2).
On Side Route, the player takes the key or leaves it behind (Histories 3–4).
These four histories produce three states at \(q\): holding a visitor pass without a key, holding a key without a pass, or holding neither (Figure~\ref{fig:state_propagation}(b)).
The first two histories share a state because they reach the same position with identical variable values.
State tuples list the values of \texttt{pass\_type} and \texttt{has\_key}, in that order.

\begin{figure*}[!t]
  \centering
  \includegraphics[width=\linewidth]{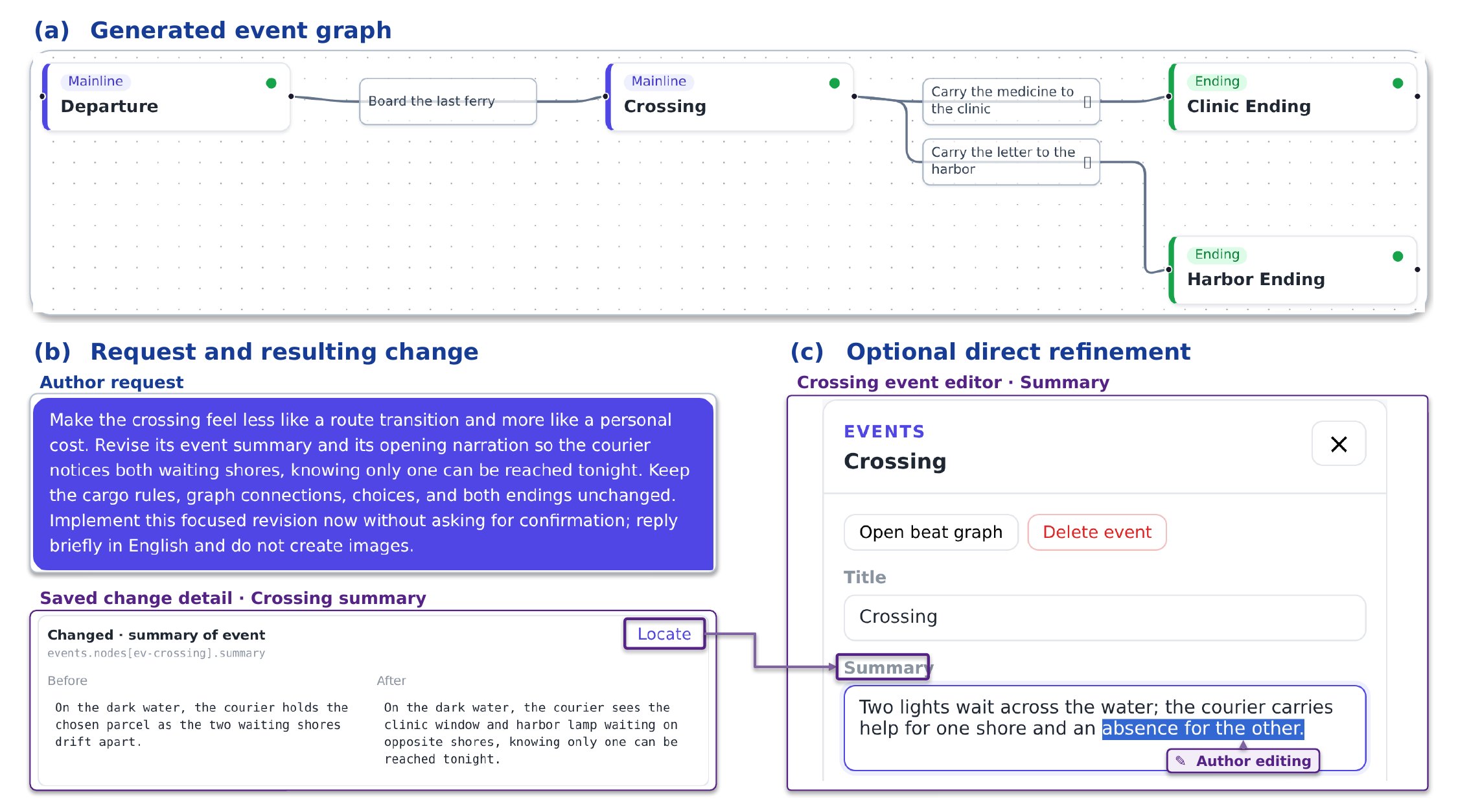}
  \caption{Delegating and refining an interactive narrative in NarrativeSteward.
  In \emph{The Last Delivery}, a courier must choose between delivering medicine to a clinic and a letter to a harbor.
  (a) The generated event graph connects Departure and Crossing to the two endings.
  (b) After reviewing Crossing's summary, the author requests revisions to the summary and opening narration to emphasize the personal cost while preserving the rules and choices.
  The screenshot shows the summary change within the changeset, including its before-and-after text.
  (c) \emph{Locate} opens the same summary in the event editor, where the author chooses to refine its wording directly, emphasizing the shore left waiting.}
  \Description{The event graph connects Departure to Crossing, then to Clinic Ending and Harbor Ending.
  A changeset excerpt compares the Crossing summary before and after the agent's revision.
  The event inspector shows the author's subsequent summary about helping one shore and leaving the other waiting; purple annotations connect Locate to Summary and mark the selected words as an author edit.}
  \label{fig:delegation_takeover}
\end{figure*}

The checkpoint event has three outgoing edges, corresponding to the Visitor, Side, and Staff gates shown in Figure~\ref{fig:state_propagation}(c).
The Visitor gate requires \texttt{pass\_type=visitor}, the Side gate requires \texttt{has\_key=true}, and the Staff gate requires \texttt{pass\_type=staff}.
Figure~\ref{fig:state_propagation}(c) checks each condition against all three reachable states.
The state \texttt{(visitor, false)}---holding a visitor pass without a key---enables the Visitor gate, while \texttt{(none, true)}---holding a key without a pass---enables the Side gate.
The state \texttt{(none, false)}, with neither a pass nor a key, leaves every gate disabled, creating a \emph{dead end}: a reachable non-ending state with no continuation.
The Staff gate is never enabled because none of the reachable states contains a staff pass.
Individual disabled cells indicate unavailable options for particular states; the dead end is identified by an entire disabled row, and the never-enabled gate by an entire disabled column.

To avoid exploring equivalent continuations repeatedly, the system merges states at the same position when their values agree on the variables needed for subsequent condition checks.
It identifies these variables by working backward from conditions across both graph levels, accounting for intervening state updates: a fixed assignment replaces a variable’s previous value, whereas an addition can carry that value forward into a later condition.
Appendix~\ref{app:technical-evaluation} details this analysis and explains why merging preserves the execution checks.

The system checks reachability and available continuations throughout the event and beat graphs.
Verification passes only when the analysis completes and confirms that every event and beat is reachable, every edge is enabled in at least one reachable state, and every reachable nonterminal state has a continuation.
For a dead end, a diagnostic trace shows how the player can reach the blocked state.
Authors can examine its associated choices, effects, and conditions when deciding how to revise the story.

\begin{figure*}[!t]
  \centering
  \includegraphics[width=\linewidth]{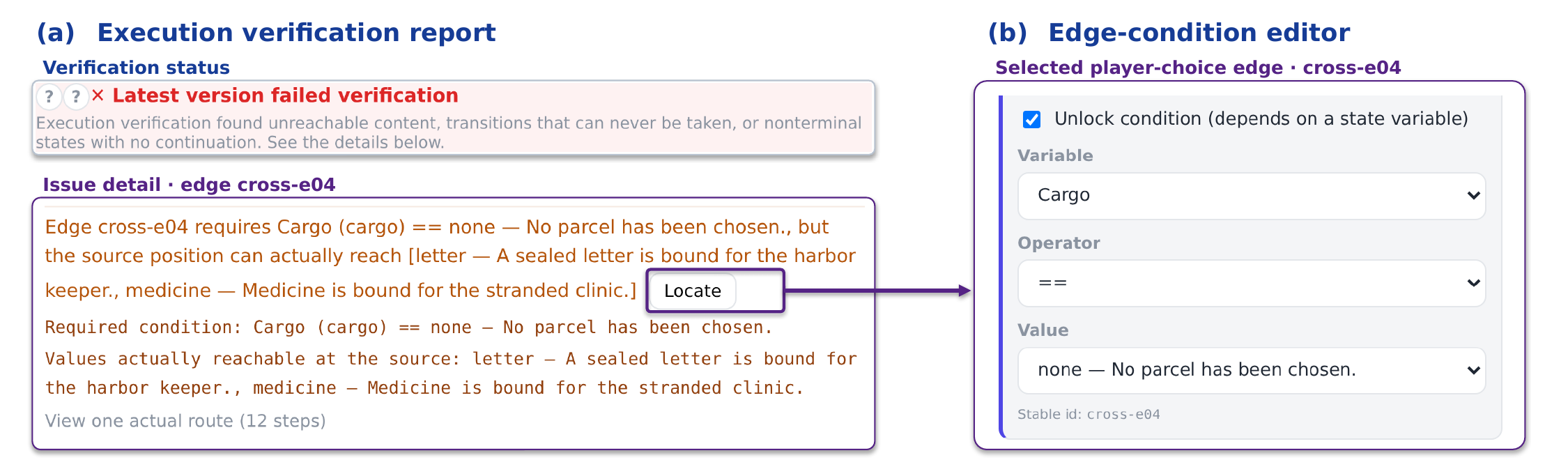}
  \caption{Diagnosing an unreachable branch in \emph{The Last Delivery}.
  (a) The author runs execution verification, which identifies a transition in Crossing that requires \texttt{cargo=none} (empty-handed).
  Earlier choices have already set \texttt{cargo} to \texttt{medicine} or \texttt{letter}, so this transition can never be taken.
  (b) The author uses \emph{Locate} to inspect the corresponding edge condition.
  To preserve the intended choice between helping the clinic and the harbor, the author subsequently removes the unreachable beat and its incoming transition.}
  \Description{A failed-verification excerpt lists the required none value and the actually reachable medicine and letter values.
  Beside it, the selected player-choice edge inspector shows the condition Cargo equals none and stable ID cross-e04.}
  \label{fig:version_bound_verification}
\end{figure*}

As authors revise the narrative, they need to know whether an earlier verification result still applies.
The system associates each report with the saved content it checked and uses a \emph{content fingerprint} to detect changes to those inputs.
When these inputs change, the interface indicates that the narrative needs to be verified again.

NarrativeSteward also supports playtesting, in which authors experience the narrative through player choices.
Playtesting applies the same beat effects and edge conditions as execution verification, allowing authors to inspect how these rules unfold along a selected route.

\begin{figure*}[!t]
  \centering
  \includegraphics[width=\linewidth]{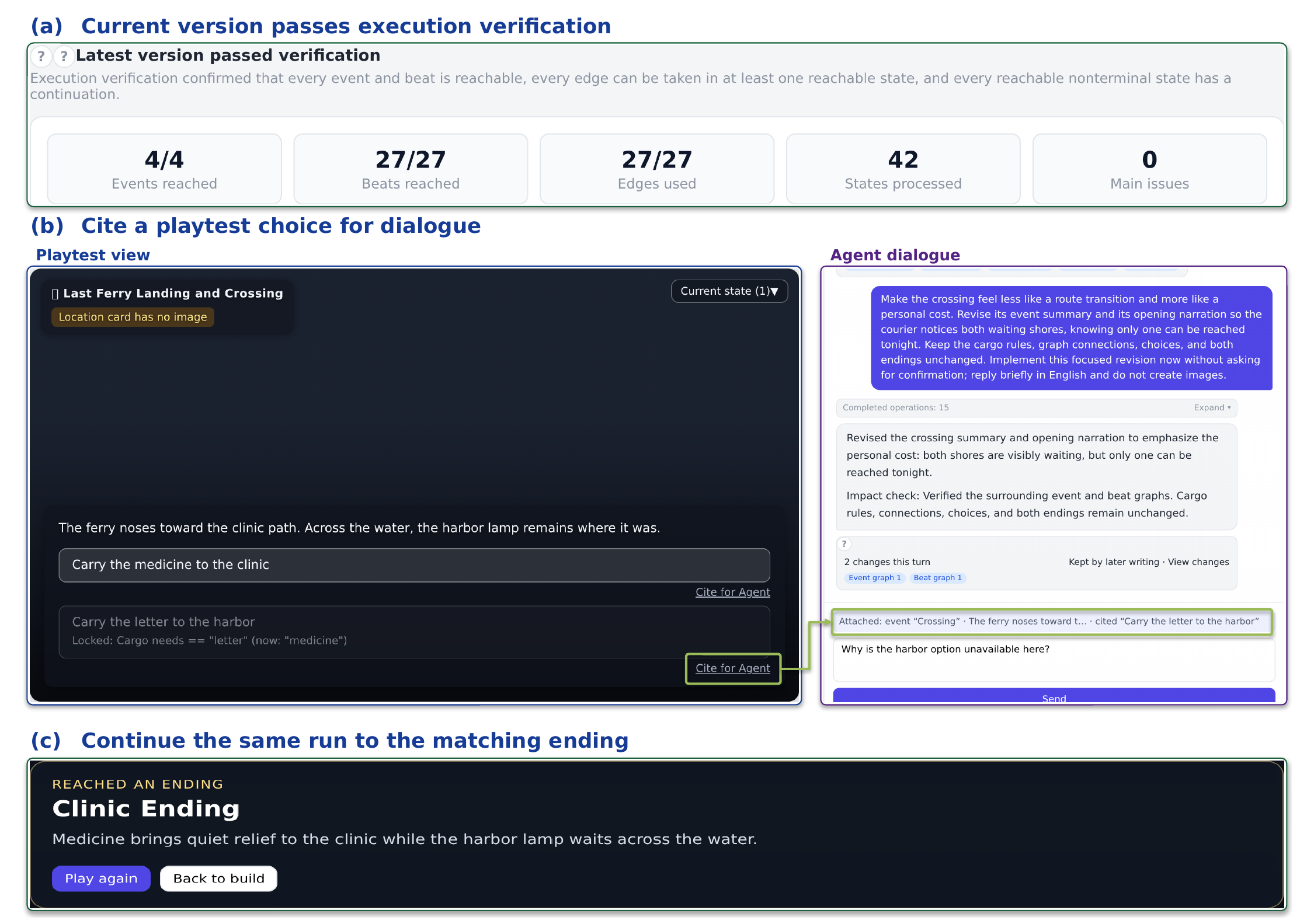}
  \caption{Verifying and playtesting \emph{The Last Delivery} after repair.
  (a) Execution verification passes after the unreachable branch is removed.
  (b) During playtesting, the author chooses to carry medicine at Departure.
  At Crossing, the clinic destination is available, while the harbor remains locked because it requires the letter.
  \emph{Cite for Agent} attaches the selected option and current playtest context to the dialogue, where the author drafts a question about the locked destination.
  (c) Continuing the same playtest reaches Clinic Ending.}
  \Description{A passed-verification summary reports complete coverage and zero main issues.
  A playtest excerpt offers the clinic and disables the harbor because cargo is medicine rather than letter; beside it, the agent dialogue panel displays attached Crossing context and the cited harbor choice while a question is drafted.
  A final excerpt shows the Clinic Ending reached by that route.}
  \label{fig:verification_playtest}
\end{figure*}

\section{Usage Scenario}
\label{sec:scenario}

Consider Maya, an author developing \emph{The Last Delivery}, a short interactive story about a courier boarding the last ferry.
The courier can carry medicine for a stranded clinic or a sealed letter for the harbor keeper, but cannot serve both destinations tonight.
Maya wants the early choice to determine the later destination and the prose to convey the cost of leaving someone waiting.
This illustrative scenario follows her from delegating an initial draft to reviewing its content, resolving an execution problem, and playtesting the result.

\subsection*{Delegating the Initial Narrative}

Maya gives the agent this premise and requests a compact story with departure, crossing, and two endings, each reachable in a different playthrough.
She authorizes continuous generation of the linked narrative artifacts.
The resulting project includes intent, an outline, world cards, an event graph, and each event's beat graph.
Maya reviews the event graph to see how Departure and Crossing connect to Clinic Ending and Harbor Ending (Figure~\ref{fig:delegation_takeover}(a)).
A shared variable, \texttt{cargo}, records the chosen parcel and determines which destination becomes available.

\subsection*{Reviewing the Narrative and Guiding Revisions}

Reading Crossing’s summary, Maya finds that it describes the courier holding the chosen parcel as the shores drift apart, but gives little sense of the cost to the people waiting there.
She asks the agent to make Crossing feel ``less like a route transition and more like a personal cost,'' revising its event summary and opening narration while preserving the cargo rules, connections, choices, and endings.
The agent updates the summary in the event graph and the narration in Crossing's beat graph.
Maya expands the changeset to compare the saved text before and after the revision (Figure~\ref{fig:delegation_takeover}(b)).
The revised summary names the waiting clinic window and harbor lamp and makes the one-destination constraint explicit.
Maya wants to place greater emphasis on what this choice means for the shore left waiting.
Using \emph{Locate}, she opens the Crossing event and its summary in the structural view.
Here, she chooses to refine the wording directly: ``Two lights wait across the water; the courier carries help for one shore and an absence for the other'' (Figure~\ref{fig:delegation_takeover}(c)).
She saves the revised summary, retaining the rules that connect the parcel choice to its destination.

\subsection*{Running Execution Verification and Choosing a Repair}

Maya requests execution verification to check whether the saved narrative's choices and rules allow its content to be reached.
The report identifies an unreachable beat and a never-enabled transition within Crossing.
The transition, ``Remain empty-handed,'' requires \texttt{cargo=none}, but only \texttt{medicine} and \texttt{letter} can reach its source (Figure~\ref{fig:version_bound_verification}(a)).
Although \texttt{cargo} starts as \texttt{none}, choosing a parcel changes its value before the courier boards the ferry.
Maya uses \emph{Locate} to inspect the transition and its condition (Figure~\ref{fig:version_bound_verification}(b)).
She could add an option to leave both parcels behind or remove the unused branch.
Adding that option would make the branch reachable, but would also let the courier avoid choosing which shore to help.
To preserve the intended dilemma of helping only one shore, she removes the unreachable beat and its incoming transition in the beat editor, keeping the original two choices and their consequences.

\subsection*{Verifying the Revision and Playtesting the Narrative}

After saving the repair, Maya runs execution verification again, and the revised narrative passes (Figure~\ref{fig:verification_playtest}(a)).
Maya then turns to playtesting to follow the early parcel choice through to its destination and narrative consequences.
She opens \emph{Playtest} and takes the medicine at Departure.
After the crossing, the clinic destination is available, while the harbor remains visible but locked because it requires the letter (Figure~\ref{fig:verification_playtest}(b)).
Using \emph{Cite for Agent}, she attaches the locked choice and its current playtest context to the agent dialogue while drafting a question about why it is unavailable.
She continues along the available route to Clinic Ending (Figure~\ref{fig:verification_playtest}(c)), where the text describes relief at the clinic and the harbor lamp still waiting across the water.
Restarting with the letter instead leads to Harbor Ending.

Across this scenario, Maya's original dilemma guides her requests, her assessment of the revised prose, and her choice of repair.
The linked artifacts let her relate these decisions to specific content and rules, while playtesting brings their consequences back into the experience of the story.

\section{Technical Evaluation}
\label{sec:tech-eval}
\newcommand{\TechDiffCases}{4}
\newcommand{\TechChangeDetails}{11}
\newcommand{\TechRecoveryCases}{4}
\newcommand{\TechUndoCases}{3}
\newcommand{\TechConflictCases}{3}
\newcommand{\TechOracleCases}{3}
\newcommand{\TechFaultPairs}{3}
\newcommand{\TechReportInvalidationCases}{5}
\newcommand{\TechReportRetentionCases}{2}
\newcommand{\TechResourceCases}{5}
\newcommand{\TechResourceRepetitions}{5}
\newcommand{\TechResourceRuns}{25}
\newcommand{\TechMaxStates}{45{,}051}
\newcommand{\TechLargestMedianMS}{59.4}
\newcommand{\TechMinMedianMS}{3.5}
\newcommand{\TechMaxMedianMS}{59.4}

The change records, recovery mechanisms, and execution diagnostics described in Section~\ref{sec:system} provide feedback and safeguards for reviewing and revising narrative artifacts.
We conducted two sets of controlled tests to evaluate whether these mechanisms accurately recorded changes, restored saved work, and produced the expected execution results.
We compared the system's behavior with predefined expected outcomes.
Table~\ref{tab:technical_summary} summarizes the results; Appendix~\ref{app:technical-evaluation} details the test cases, measurement procedure, and resource use.

\subsection{Change Records and Recovery}
\label{sec:change-recovery-evaluation}

We first tested whether changesets identified the actual changes and located the affected narrative objects.
\TechDiffCases{} cases covered local edits, batches within a beat graph, cross-layer additions, and text-and-asset changes, comprising \TechChangeDetails{} expected change details.
The system identified all \TechChangeDetails{} details, reported no additional details, and correctly located every affected object.

We then tested recovery from failed or stopped turns and whole-turn undo.
Failures and stops were introduced into the implementation's turn-finalization process, which saves or restores the turn's changes.
All \TechRecoveryCases{} failure or stop cases restored the project to its state before the turn without producing a successful changeset.
In \TechUndoCases{} undo cases without conflicts, the system restored prior content.
In \TechConflictCases{} cases with subsequent conflicting edits or revision changes, it rejected undo while preserving the saved work, without partially reverting the turn.
These tests check the recovery and undo behavior described in Section~\ref{sec:inspectable-edits}.

\begin{table*}[!t]
  \centering\small
  \caption{Controlled evaluation of NarrativeSteward's change records, recovery, and execution verification.
Fractions indicate checks that matched predefined expectations out of the tested items or runs.
The first three rows assess \TechChangeDetails{} individual change details across \TechDiffCases{} editing cases.
Recovery and undo checks assess restoration of prior work and protection of subsequent edits.
Fault tests use paired faulty and corrected projects.
Report invalidation and retention assess whether previous verification reports remain applicable after changes to checked content or to descriptions and image paths, respectively.
Completion counts runs that finished with either a pass or a fault diagnosis.}
  \label{tab:technical_summary}
\begin{tabular}{@{}p{0.36\linewidth}p{0.28\linewidth}p{0.29\linewidth}@{}}
\toprule
Check & Test unit & Result \\
\midrule
Change detection & Expected change details & 11/11 detected \\
Change precision & Reported change details & 11/11 correct \\
Change localization & Locatable change details & 11/11 matched \\
Failure / stop recovery & Failure or stop cases & 4/4 restored \\
Whole-turn undo & Cases without conflicts & 3/3 restored \\
Undo conflict protection & Cases with conflicts & 3/3 protected \\
\midrule
Execution semantics & Hand-computed cases & 3/3 matched \\
Fault localization & Faulty projects & 3/3 located \\
Verification of corrected projects & Corresponding corrected projects & 3/3 passed \\
Report invalidation & Edits to checked inputs & 5/5 invalidated \\
Report retention & Description / image-path edits & 2/2 retained \\
Verification completion & Runs: 5 cases $\times$ 5 repetitions & 25/25 completed \\
\bottomrule
\end{tabular}

\end{table*}

\subsection{Execution Verification}
\label{sec:execution-evaluation}

We tested whether execution verification followed the narrative's state rules, identified faults, and maintained the applicability of previous reports.
The system matched the expected results in \TechOracleCases{} hand-computed cases covering linear flow, updates within bounded variable ranges, and overwriting variable values after branches converge.
We prepared \TechFaultPairs{} pairs of faulty and corrected projects, covering an unreachable event, a never-enabled edge, and a reachable dead-end state.
The system identified the expected fault and its location in each faulty version, and all \TechFaultPairs{} corrected versions passed execution verification.

To test whether verification reports remained applicable after revisions, we prepared \TechReportInvalidationCases{} edits to conditions, state updates, graph connections, or referenced character identities---all of which form part of the content checked by execution verification.
We also prepared \TechReportRetentionCases{} edits to world-card descriptions or image paths, which do not affect these checks and therefore allow the previous report to remain applicable.
The system correctly invalidated the reports in all \TechReportInvalidationCases{} cases requiring re-verification and retained them in the other \TechReportRetentionCases{}.

We measured execution verification time on \TechResourceCases{} test cases covering combinations of narrative choices and state updates, as well as variations in variable ranges, the number of simultaneously relevant variables, branch width, and the distance to a fault requiring diagnostic-route reconstruction.
Each case was measured \TechResourceRepetitions{} times, and all \TechResourceRuns{} runs completed.
Across the \TechResourceCases{} cases, median verification times ranged from \TechMinMedianMS{} to \TechMaxMedianMS{}\,ms, with the longest median observed in the case that explored \TechMaxStates{} execution states.

\section{User Study}
\label{sec:study}

We conducted a within-subjects study combining questionnaires, interview accounts, and system-use records to examine delegation experience and effort (RQ1), understanding and continued refinement (RQ2), and feedback-informed judgments and revision decisions (RQ3).

\subsection{Participants}
\label{sec:participants}

We recruited 12 participants (7 men and 5 women), aged 22--37 years ($M=27.5$, $SD=5.1$).
Their reported creative experience ranged from none ($n=2$) to less than one year ($n=4$), one to less than four years ($n=5$), and seven or more years ($n=1$).
Backgrounds included writing and game or narrative design; 8 reported student or researcher roles related to interactive narrative, HCI, or generative AI.
Three participants had not used interactive-narrative, story-graph, or game-editing tools in the preceding year, while three used them at least weekly.
All had previously used their selected general-purpose agent to complete a multi-file project that could be previewed or run; 10 used that agent at least weekly and 2 used it one to three days per month in the preceding six months.

Participation was voluntary and uncompensated, with consent obtained for participation and data use.
Findings are reported using anonymized participant identifiers.

\subsection{Study Design and Procedure}
\label{sec:procedure}

To examine how NarrativeSteward supports authors in understanding and guiding interactive narratives while delegating implementation, we compared it with participants' familiar general-purpose agents in a within-subjects design.
We chose this baseline because it supports the implementation of runnable projects and allows participants to draw on their existing tool experience.
A comparison with a primarily manual narrative editor would instead change how much implementation could be delegated, shifting the focus of the evaluation.
The study evaluates authors' experiences with the two authoring environments, including their available models and tools.

In the NarrativeSteward condition, participants could freely choose which features to use and when to use them, including dialogue, structural views, direct editing, playtesting, and execution verification.
The system used 
\texttt{gpt-5.5} with the default \texttt{medium} reasoning effort%
 for all participants; further implementation details appear in Appendix~\ref{app:implementation}.
For the baseline, 8 participants selected Cursor and 4 selected Codex.

They could use native capabilities such as file editing, terminal access, change review, undo, and preview, and could edit code themselves.
Participants could choose their model and reasoning settings, reported the displayed configuration where available, and were asked to keep it fixed throughout the task.
In the baseline condition, six participants reported using Grok, three GPT, and one Claude; the remaining two did not specify their model.

Each participant created a different story from blank content with each tool, and we counterbalanced the order of the two conditions.
Both conditions asked participants to create a browser-runnable interactive story with consequential choices, delayed effects of earlier decisions, and multiple endings.
Participants chose their own topics, which could share a broad genre across conditions.
We offered a starting scope of three to five main story stages, six to ten meaningful decision points, and two to three reachable endings.
We also suggested that at least two early decisions affect later content, available actions, or endings, and that a complete route take approximately 8--12 minutes to play.
Participants could adapt this starting scope and develop larger stories.

Participants read the study overview and completed a background questionnaire before starting their first assigned condition.
Immediately before the NarrativeSteward task, they read usage instructions and could consult in-system guidance; participants starting with the baseline did not receive this introduction beforehand.
Researchers were available to answer questions during the tasks.
Participants decided when to submit, without a fixed creation time limit.
NarrativeSteward projects were submitted through the study platform; baseline submissions contained a static web project and launch instructions.
Researchers manually checked the submitted projects and confirmed that all projects in both conditions were runnable.
Appendix~\ref{app:story-scale} summarizes the topics and scale of the completed narratives.
Participants completed a condition-specific questionnaire after each authoring task.

After completing both tasks, participants provided cross-tool comparisons and participated in interviews about their experiences and decisions, with interview responses documented in notes.

\subsection{Data Collection and Analysis}
\label{sec:measures}
\label{sec:analysis}

We combined questionnaires, system-use records, and interview accounts to examine authors' experiences and how they understood and guided their work during delegated creation.
Questionnaires assessed both authoring conditions using seven-point ratings with question-specific anchors.
Thirteen measures addressed delegation experience (RQ1), understanding and continued refinement (RQ2), and feedback-informed judgments and revision decisions (RQ3).
Additional questions examined desired but unfinished revisions and reasons for stopping, as well as tool preferences across authoring activities and overall.
For NarrativeSteward, participants reported their use of seven authoring features and rated the helpfulness of those they had used on a seven-point scale (1 = not at all helpful, 4 = somewhat helpful, 7 = extremely helpful).
They also completed the 10-item System Usability Scale (SUS), using its published Simplified Chinese wording and standard five-point response scale \cite{brooke1996sus,wang2020chinesesus}.
SUS responses were converted to a score from 0 to 100 using standard scoring.
Questionnaire measures, anchors, and applicability conditions are described in Appendix~\ref{app:questionnaire}.

We analyzed each rating item separately, using eligible numeric responses from both conditions for paired comparisons and reporting valid paired sample sizes.
We summarized rating distributions and paired differences and used two-sided Wilcoxon signed-rank tests with paired rank-biserial effect sizes.
Holm--Bonferroni correction was applied separately within RQ1 (four measures), RQ2 (five), and RQ3 (four).
Responses indicating no relevant experience or inability to judge were excluded from numeric comparisons rather than assigned midpoint scores.
Other responses were summarized descriptively: feature helpfulness among self-reported users providing numeric ratings and stopping reasons among those reporting desired but unfinished revisions.
Detailed statistical procedures and supplementary results appear in Appendix~\ref{app:paired_comparisons}.

NarrativeSteward recorded system-use events, changesets, and verification metadata during authoring, covering actions such as reviewing changes, saving direct edits, undoing turns, running verification, and playtesting.
Questionnaires and interviews supplied accounts of general-purpose agent use.

We used thematic analysis to examine participants' experiences of forming requests, inspecting the work, and deciding what to revise or check next \cite{braun2006using}.
The interview guide appears in Appendix~\ref{app:interview-guide}.
Two researchers reviewed the interview notes, coded relevant passages, and discussed similarities and differences across participants to develop themes addressing the research questions.
The analysis considered authors' aims, the information they consulted, their interpretation of that information, and subsequent decisions.
The researchers revisited the material to check and refine these themes, considering both supporting and contrasting experiences.
Interview notes and written accounts are presented as paraphrases attributed to anonymized participants.

\section{User Study Findings}
\label{sec:findings}

\begin{figure*}[!t]
  \centering
  \includegraphics[width=\linewidth]{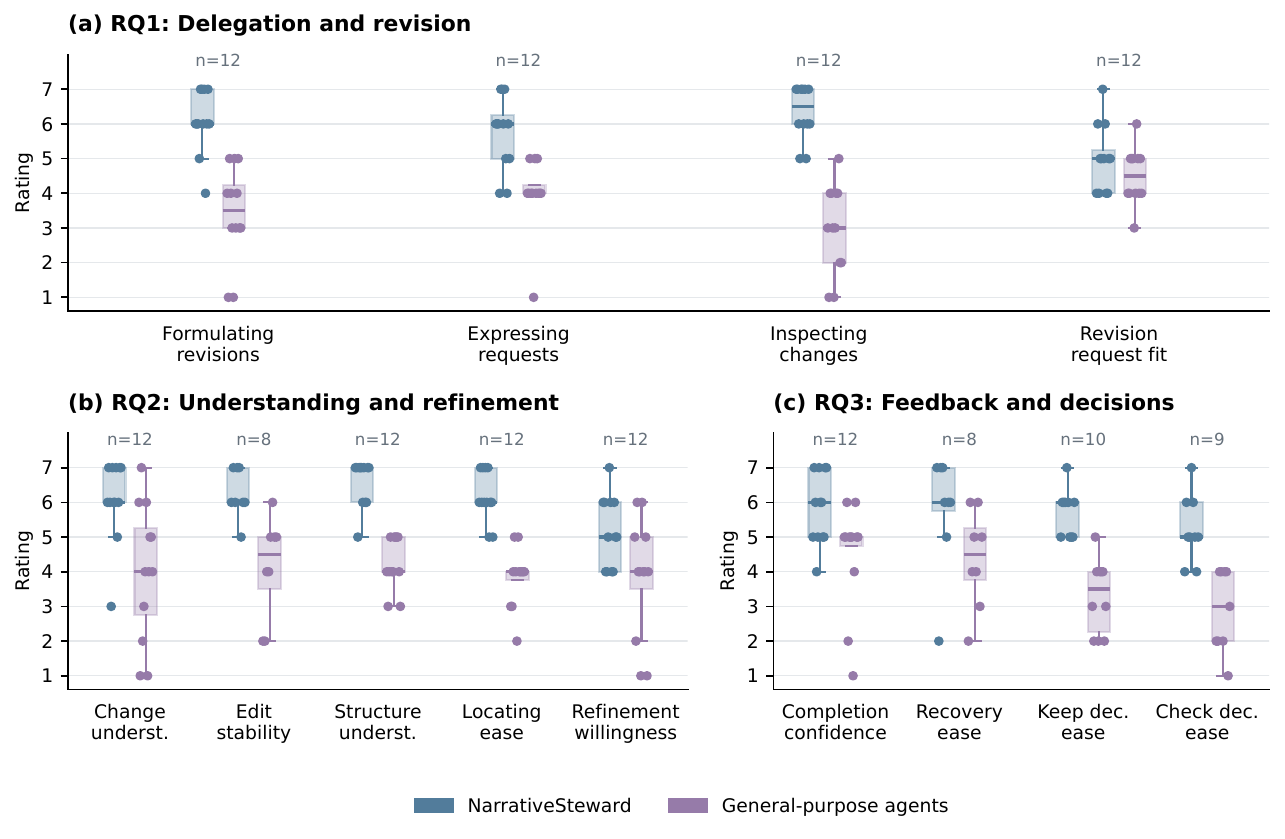}
  \caption{Self-reported authoring experience with NarrativeSteward and participants' selected general-purpose agents, rated on seven-point scales.
  Panels cover (a) delegation and revision, (b) understanding and refinement, and (c) feedback and decisions.
  Higher scores indicate more favorable experiences.
  Each item includes the same eligible participants in both conditions; $n$ gives the number of pairs.
  Boxes show the interquartile range and median, whiskers extend to observations within 1.5 interquartile ranges of the box, and dots show individual ratings with horizontal offsets to separate responses.
  ``Underst.'' abbreviates understanding; ``Keep dec. ease'' and ``Check dec. ease'' refer to deciding whether to retain a modification or check further.}
  \Description{Three panels compare thirteen experience measures for NarrativeSteward and general-purpose agents using vertical boxplots and individual scores. RQ1 spans the top row, with RQ2 and RQ3 below. Numeric-pair denominators vary with reported experience.}
  \label{fig:study_experience}
\end{figure*}

We present the findings by research question, combining comparisons of authoring experience (Figure~\ref{fig:study_experience} and Table~\ref{tab:study_experience_summary}) with participants' accounts of forming requests, understanding the work, and making revision decisions.

\begin{table*}[!t]
  \caption{Paired comparisons of self-reported authoring experience on seven-point scales.
  Higher scores indicate more positive ratings on each measure.
  NS denotes NarrativeSteward; Agent denotes participants' selected general-purpose agents.
  Paired $n$ counts eligible responses from the same participants in both conditions; $M$ and $SD$ give their mean and sample standard deviation.
  The paired rank-biserial effect size $r_{rb}$ is positive when NS ratings are higher.
  Adjusted $p$-values are from two-sided Wilcoxon signed-rank tests with Holm--Bonferroni correction within RQ1 (four measures), RQ2 (five), and RQ3 (four).}
  \label{tab:study_experience_summary}
  \centering\small
  \setlength{\tabcolsep}{4pt}
  \begin{tabular}{@{}llrccccrr@{}}
\toprule
& & & \multicolumn{2}{c}{NarrativeSteward} & \multicolumn{2}{c}{Agent} & & \\
\cmidrule(lr){4-5}\cmidrule(lr){6-7}
RQ & Measure & Paired $n$ & $M$ & $SD$ & $M$ & $SD$ & $r_{rb}$ & Adj. $p$ \\
\midrule
\multirow{4}{*}{RQ1} & Formulating revisions & 12 & 6.08 & 0.90 & 3.42 & 1.38 & 1.00 & 0.003 \\
 & Expressing requests & 12 & 5.75 & 1.06 & 4.00 & 1.04 & 1.00 & 0.008 \\
 & Inspecting changes & 12 & 6.33 & 0.78 & 3.00 & 1.28 & 1.00 & 0.002 \\
 & Revision request fit & 12 & 5.00 & 0.95 & 4.50 & 0.80 & 0.52 & 0.375 \\
\midrule
\multirow{5}{*}{RQ2} & Change understanding & 12 & 6.08 & 1.16 & 4.00 & 1.95 & 0.91 & 0.029 \\
 & Edit stability & 8 & 6.25 & 0.71 & 4.12 & 1.46 & 0.89 & 0.062 \\
 & Structure understanding & 12 & 6.58 & 0.67 & 4.25 & 0.75 & 1.00 & 0.002 \\
 & Locating ease & 12 & 6.17 & 0.72 & 3.83 & 0.83 & 1.00 & 0.002 \\
 & Refinement willingness & 12 & 5.17 & 1.03 & 3.83 & 1.70 & 0.72 & 0.094 \\
\midrule
\multirow{4}{*}{RQ3} & Completion confidence & 12 & 5.83 & 1.03 & 4.50 & 1.51 & 0.91 & 0.031 \\
 & Recovery ease & 8 & 5.75 & 1.67 & 4.38 & 1.41 & 0.53 & 0.211 \\
 & Keep decision ease & 10 & 5.70 & 0.67 & 3.30 & 1.06 & 1.00 & 0.008 \\
 & Check decision ease & 9 & 5.22 & 0.97 & 2.89 & 1.17 & 1.00 & 0.012 \\
\bottomrule
\end{tabular}

\end{table*}

\subsection{RQ1: Delegation Experience and Effort}
\label{sec:findings-rq1}

\paragraph{Formulating requests and inspecting changes were easier with NarrativeSteward.}
Participants rated identifying desired revisions, expressing revision requests, and inspecting changes more favorably with NarrativeSteward than with their general-purpose agents.
Across the \StudyFormulateN{} paired responses, median ratings for NarrativeSteward versus general-purpose agents were \StudyFormulateNS{} versus \StudyFormulateAgent{} for formulating revisions, \StudyExpressNS{} versus \StudyExpressAgent{} for expressing requests, and \StudyInspectChangesNS{} versus \StudyInspectChangesAgent{} for inspecting changes; all three differences were significant after within-RQ correction (adjusted $p = \StudyFormulateP{}$, $\StudyExpressP{}$, and $\StudyInspectChangesP{}$, respectively).

P6 found it difficult to identify revision targets in their general-purpose agent's textual output; a requested mind map remained difficult to use, and they eventually stopped revising.
In NarrativeSteward, P6 described moving between story details and an overview of the creative intent.
P2 described inspecting individual parts of the work in NarrativeSteward, whereas checking changes with their general-purpose agent required rereading the output and following branches.
P8 similarly valued seeing intermediate content and checking requirements against individual parts.
These accounts connect the ease of forming and reviewing requests with access to specific content and its broader context.

The fit between revision requests and their results did not differ significantly after correction (medians \StudyRequestFitNS{} versus \StudyRequestFitAgent{}; adjusted $p = \StudyRequestFitP{}$).
P3 found changing a setting in NarrativeSteward straightforward, but related plot content required additional requests to update.
P10 described several rounds of clarification when the agent overused a supplied reference and repeatedly exposed a clue intended to remain subtle.
Both accounts show the additional coordination involved in carrying a requirement through related content or achieving its intended narrative effect.

\paragraph{Review effort and willingness to invest further shaped continued revision.}
All \StudyCompleteN{} participants requested revisions in both conditions, while two reported desired but unfinished revisions with NarrativeSteward and five with their general-purpose agent (Table~\ref{tab:study_revision_experience}).
Of the latter five, four selected anticipated effort to inspect results, two selected anticipated communication effort, and two selected unwillingness to invest further; participants could select up to two reasons.
For NarrativeSteward, one respondent selected limited time or other activities, and one selected unwillingness to invest further.

\begin{table*}[!t]
  \centering\small
  \caption{Revision requests, unfinished revisions, and reasons for stopping.
  Counts refer to participants using NarrativeSteward (NS) or their selected general-purpose agent (Agent).
  The first two groups summarize whether participants requested revisions during the task and whether desired revisions remained unfinished at its end (\StudyCompleteN{} participants per condition).
  Stopping reasons apply only to participants with unfinished revisions (NS: \StudyUnfinishedNS{}; Agent: \StudyUnfinishedAgent{}).
  Each could select up to two reasons, so reason counts can overlap.}
  \label{tab:study_revision_experience}
  \begin{tabular}{@{}llrr@{}}
\toprule
Category & Response & NS & Agent \\
\midrule
\multirow{2}{*}{\shortstack[l]{Revision\\requests}} & Requested a revision & 12 & 12 \\
 & Considered revisions but made no request & 0 & 0 \\
\midrule
\multirow{2}{*}{\shortstack[l]{Unfinished\\revisions}} & Had unfinished revisions & 2 & 5 \\
 & Had no unfinished revisions & 10 & 7 \\
\midrule
\multirow{4}{*}{\shortstack[l]{Reasons for\\stopping}} & Did not want to invest further & 1 & 2 \\
 & Limited time / other activities & 1 & 0 \\
 & Expected effort to inspect results & 0 & 4 \\
 & Expected communication effort & 0 & 2 \\
\bottomrule
\end{tabular}

\end{table*}

P11 found branch checking with their general-purpose agent cumbersome and was reluctant to keep polishing the output.
P6 distinguished enjoyable effort spent developing the story from fatigue caused by rereading successive complete outputs.
Other participants stopped for different reasons: P8 had achieved the intended direction in both works, whereas P10 still wanted to improve branches and game feedback after playtesting but did not want to invest further in either condition.
These experiences show how the work of checking and communicating revisions, together with authors' satisfaction and available effort, shaped decisions about continuing creation.

\subsection{RQ2: Understanding and Continuing to Refine the Work}
\label{sec:findings-rq2}

\paragraph{Structural review helped authors develop subsequent requests.}
All \StudyOverviewN{} participants rated understanding the overall structure and locating content higher with NarrativeSteward.
Median ratings were \StudyOverviewNS{} versus \StudyOverviewAgent{} for structural understanding and \StudyLocateNS{} versus \StudyLocateAgent{} for locating ease (both adjusted $p = \StudyOverviewP{}$).
Understanding agent changes also received higher ratings (medians \StudyUnderstandingNS{} versus \StudyUnderstandingAgent{}; adjusted $p = \StudyUnderstandingP{}$).
Local-edit stability had medians of \StudyStabilityNS{} versus \StudyStabilityAgent{}, with eight eligible pairs; the difference was not significant after correction (adjusted $p = \StudyStabilityP{}$).

P1's experience illustrates how reviewing the generated structure helped identify a target for further implementation.
After the agent generated an initial event graph, P1 noticed that earlier branches had little consequence and that different endings depended mainly on the final choice.
P1 raised this concern with the agent, which analyzed the issue and proposed a way to restructure the branches.
P1 accepted the proposal and asked the agent to carry out the revision.
P1 thus used structural review to identify what needed to change and the agent's proposal to decide how to proceed with the revision.

Authors also moved between related parts of the narrative when reviewing changes and deciding how to revise the work.
P2 inspected an added character and the revised storyline in separate content views.
P6 edited a character name directly and then asked the agent to update related content.
P10 combined delegated revision with direct wording edits and associated the visible branches and updated content with a clearer understanding of what had changed.
These accounts show authors moving between reviewing the work and selecting how to pursue a revision.

\begin{figure*}[!t]
  \centering
  \includegraphics[width=\linewidth]{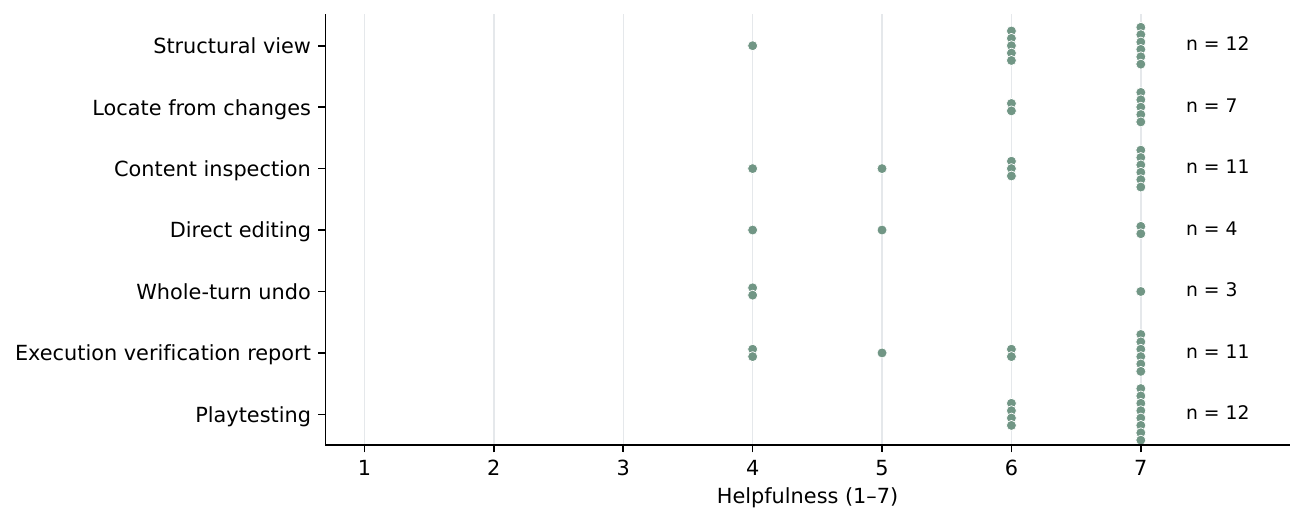}
  \caption{Helpfulness ratings for seven NarrativeSteward features, provided by participants who reported using each feature.
  Each dot represents one valid rating; $n$ gives the number of ratings for that feature, and vertical offsets separate overlapping points.
  Ratings range from 1 (not at all helpful), through 4 (somewhat helpful), to 7 (extremely helpful).}
  \Description{Seven rows show individual helpfulness ratings for the structural view, locating changes, inspecting content, direct editing, whole-turn undo, execution verification reports, and playtesting. Scores run from 1 to 7; each row displays its own valid rating count.}
  \label{fig:study_system_use}
\end{figure*}

Participants' helpfulness ratings describe the support they found in these features (Figure~\ref{fig:study_system_use}).
On the seven-point scale, the structural view and content inspection received median ratings of \StudyGraphsMedian{} ($n = \StudyGraphsRatedN{}$) and \StudyContentMedian{} ($n = \StudyContentRatedN{}$), respectively.
Direct editing received a median of \StudyDirectMedian{} ($n = \StudyDirectRatedN{}$), and whole-turn undo a median of \StudyUndoMedian{} ($n = \StudyUndoRatedN{}$).
Together with the participant accounts, these ratings indicate that authors valued support for examining the work while also drawing on local editing and recovery when needed.

\paragraph{Refinement preferences reflected the task and the desired flexibility.}
Overall, seven participants preferred NarrativeSteward, one preferred their general-purpose agent, one reported no preference, and three selected insufficient experience.
Tool-preference counts combine ``probably'' and ``definitely'' responses for each tool.
NarrativeSteward's System Usability Scale score on the 0--100 scale had a median of \StudySUSMedian{} (range \StudySUSMin{}--\StudySUSMax{}).

Willingness to continue refining was higher with NarrativeSteward in seven pairs, equal in four, and lower in one, with medians of \StudyRefinementNS{} versus \StudyRefinementAgent{}; the difference was not significant after correction (adjusted $p = \StudyRefinementP{}$).
P6 associated structural views and support for developing ideas with wanting to spend more time on the work.
P12 saw branches and cards as suitable for detailed refinement when time and interest permitted.
In tool choices for different authoring activities (Table~\ref{tab:study_stages}), eight participants selected NarrativeSteward for polishing and eleven for checking or delivery.
P8 preferred combining brainstorming with a general-purpose agent and authoring with NarrativeSteward.

\begin{table*}[!t]
  \centering\small
  \caption{Participants' preferred tools for four interactive-narrative authoring activities.
  Each row counts one choice per participant ($N=\StudyCompleteN{}$).
  NS denotes NarrativeSteward; Agent denotes participants' selected general-purpose agents.
  Both means using the tools in combination, Either means either tool would suffice, Neither means neither would suit the activity, and Unsure indicates insufficient experience to choose.}
  \label{tab:study_stages}
  \begin{tabular}{@{}lrrrrrr@{}}
\toprule
Authoring activity & NS & Agent & Both & Either & Neither & Unsure \\
\midrule
Ideation & 6 & 1 & 2 & 3 & 0 & 0 \\
First playable version & 6 & 3 & 1 & 1 & 0 & 1 \\
Polishing & 8 & 2 & 2 & 0 & 0 & 0 \\
Checking / delivery & 11 & 1 & 0 & 0 & 0 & 0 \\
\bottomrule
\end{tabular}

\end{table*}

Participants also valued options outside NarrativeSteward's current organization.
P3 preferred their general-purpose agent's ability to present the story in a form adapted to their needs.
P4 valued modular organization but favored a general-purpose agent for a longer project where they could develop additional tools.
P7 wanted earlier previews and more flexible illustration generation.
P9 appreciated automated implementation but found that receiving a large amount of content at once created pressure to confirm too much.
For ideation, P9 preferred their familiar agent's knowledge base and conversational habits, while finding NarrativeSteward attractive for interactive implementation.
These accounts connect tool preferences with the kind of work authors wanted to pursue and when they wanted to inspect its results.

\subsection{RQ3: Feedback, Judgment, and Revision Decisions}
\label{sec:findings-rq3}

\paragraph{Authors found it easier to judge whether to retain changes or check further.}
Ease of deciding whether to retain a modification had medians of \StudyKeepNS{} with NarrativeSteward and \StudyKeepAgent{} with general-purpose agents; ease of deciding whether more checking was needed had medians of \StudyCheckNS{} versus \StudyCheckAgent{}.
All ten eligible pairs favored NarrativeSteward for the former and all nine for the latter (adjusted $p = \StudyKeepP{}$ and $\StudyCheckP{}$, respectively).
Task-completion confidence was also higher with NarrativeSteward across \StudyConfidenceN{} pairs (medians \StudyConfidenceNS{} versus \StudyConfidenceAgent{}; adjusted $p = \StudyConfidenceP{}$).
Recovery ease had medians of \StudyRecoveryNS{} versus \StudyRecoveryAgent{} across eight pairs, with no significant difference after correction (adjusted $p = \StudyRecoveryP{}$).

Participants drew on content, changes, and playtesting when making these judgments.
P12 briefly examined which kinds of content had changed in NarrativeSteward, while consulting their general-purpose agent's summary before trying its demo.
P4 assessed the opening through play and distinguished an executable flow from prose and responses that felt appropriate.
P10 reported awareness without use of execution verification and relied on playtesting to judge runtime behavior.
These accounts show how authors assessed both what had been implemented and how it felt when experienced as a story.

\paragraph{Diagnostics and playtesting informed further revision and checking.}
Execution verification reports and playtesting each received a median helpfulness rating of \StudyVerificationHelpMedian{} on the seven-point scale ($n = \StudyVerificationHelpRatedN{}$ and $n = \StudyPlaytestHelpRatedN{}$, respectively; Figure~\ref{fig:study_system_use}).

P5 described paying little attention to file-level changes with their general-purpose agent because locating the relevant narrative content was inconvenient.
In NarrativeSteward, P5 used both execution diagnoses and the agent's analysis of story consistency to direct successive revisions.
P5 first asked the agent to generate the complete worldbuilding and interactive narrative, then inspected the characters and locations and manually changed some names.
P5 asked the agent to identify and update other text affected by the renaming.
After running execution verification and finding problems, P5 delegated their repair to the agent.
P5 next asked the agent to examine the whole story for semantic inconsistencies, reviewed its feedback, and asked it to revise the story.
P5 then ran execution verification again on the revised work; when it identified execution problems, P5 asked the agent to repair them.
This sequence shows P5 directing successive implementation tasks through content inspection, agent-generated semantic feedback, and system-generated execution diagnoses.

P11 also described locating a problem from a diagnostic and asking the agent to repair it.
P8 reran execution verification after local revisions and used the outcome to decide whether to retain the change.
With their general-purpose agent, P2 found that an announcement of completed revision did not always match the subsequent playthrough, prompting another check.
These episodes connect feedback with concrete revision targets and decisions to examine the work again.

\section{Discussion}
\label{sec:discussion}

\subsection{Structural Review Helps Authors Develop Creative Direction}

When an autonomous agent develops a narrative from a broad request, many choices and relationships take shape before the author has specified them individually.
Structural review gives authors a way to examine these emerging decisions and turn their creative aims into concrete revision requests.
An author may want earlier choices to matter without having specified which choices should affect which later events or endings.
Examining a generated version makes those relationships available for consideration and can reveal where the author wants a different outcome.
P1's concern about inconsequential branches illustrates this process: inspecting the event graph prompted a discussion in which a revision direction took shape.
The generated work thus provided material for developing the next requirement.
When substantial implementation is delegated, structural review can serve both to assess the current work and to elaborate what the author wants it to become.
Supporting this process gives authors opportunities to develop their direction as the narrative takes form.

Turning creative aims into specific revision requests involves connecting the author's concern with the relevant narrative content.
DirectGPT grounds requests in visible objects, and CoNoder connects graph editing with conversational assistance \cite{masson2024direct,li2026exploring}.
In interactive narratives, judging whether a choice matters can involve its local wording, the event branches it affects, and the endings those branches make possible.
Linked narrative artifacts let authors consider these aspects together while discussing and revising the same work.
The design value lies in maintaining continuity between the concern an author identifies, the content an agent changes, and the result the author subsequently examines.
Preserving object references and access to related content across dialogue, implementation, and review can support this continuity.
A concern discovered during inspection can become a request involving the relevant narrative objects, and the resulting changes can then provide a basis for assessing that concern again.
Through this process, authors can continue to develop and guide the narrative while delegating its implementation.

\subsection{Review Effort and Flexibility in Authoring Shape Continued Delegation}

The anticipated effort of reviewing another version can influence whether authors pursue a desired revision.
Some participants left revisions unfinished with their general-purpose agents because they expected further effort to inspect results or communicate changes.
Research on AI-assisted programming similarly identifies reviewing and revising suggestions as substantial parts of interaction \cite{mozannar2024reading}.
P6's distinction between enjoyable story development and tiring rereading adds a consideration for creative work: effort can feel worthwhile when it advances a creative aim, yet discouraging when it repeatedly goes into understanding what has changed.
This suggests that supporting continued delegation involves helping authors direct their attention toward the decisions they want to make.
Making those decisions easier to reach can support further refinement when authors have both a desired improvement and a willingness to invest in it.

How review is organized can affect where that attention goes.
Requests for earlier previews and the pressure of facing large amounts of generated content suggest that authors need ways to choose when to inspect the work and how much to examine at once.
A project-wide view provides access to the whole narrative, but the question an author is currently considering may concern only part of it.
An important design consideration is therefore to make focused review possible while keeping related content accessible.
Future systems could let authors expand changes around a selected concern, inspect intermediate results, or continue delegating before reviewing a larger revision.
Such choices would help authors match review to the scope of their current question and the effort they are prepared to spend.

Whether authors continue working with an agent also depends on how well the authoring environment accommodates their preferred ways of creating.
Participants' preferences for different tools across authoring activities, and their interest in alternative presentations or additional tools, suggest that useful structural support can coexist with a need to change how the work is approached.
Research on screenwriters likewise shows that creators develop their own strategies for involving AI \cite{tang2026how}.
For dedicated authoring environments, this motivates combining an immediately usable organization of the story with opportunities to adapt its presentation and connect it to other tools.
Additional views or the transfer of selected artifacts could support those choices while preserving access to relevant narrative content and relationships.
The broader implication is to let authors adjust how they inspect and develop the work as their creative needs change.

\subsection{Feedback Connects Execution Evidence with Creative Judgment}

Feedback helped authors decide what to revise and what to ask the agent to examine next.
P5's use of execution verification and agent-led consistency review illustrates how authors can draw on different assessments of the same work to organize successive requests.
P4's distinction between an executable flow and appropriate prose and responses further shows why judging a work involves more than its ability to run.
Change records identify what was modified, execution diagnoses expose problems in the encoded choices and state rules, agent analysis offers interpretations of story consistency, and playtesting lets authors experience a selected route.
These sources offer different grounds for judging a revision, which authors relate to the creative question they are trying to resolve.

Execution verification can identify a problem, but deciding how to revise the story requires considering the author's creative goals.
An unreachable branch could be made accessible by changing a condition or removed because it no longer serves the intended story.
Either approach could resolve that particular reachability issue, but each would create different possibilities for the player.
The author therefore needs to consider what the branch contributes to the story when choosing a response.
Narrative analysis tools such as DendryScope use executable representations to help designers investigate possible play \cite{otto2023dendryscope}.
In delegated authoring, connecting such evidence to the affected content can also support discussion of alternative repairs with an agent.
The diagnosis then informs a creative choice about what to implement next.

This relationship suggests presenting feedback with enough context for authors to judge its implications.
For a blocked continuation, that context includes the relevant conditions, the content that cannot be reached, and its relation to the surrounding story.
An agent can help explain possible changes and implement the author's chosen response.
After the agent implements a revision, the author can decide which aspects of the new version need checking.
A revision aimed at story consistency may call for another execution check, while a version that passes verification may still warrant reading or playtesting to assess its narrative effect.
Connecting feedback to the affected content and subsequent requests can therefore support continued guidance across revisions.

\subsection{Limitations}
\label{sec:limitations}

Our study has several limitations that suggest directions for future research.

First, the study compares two complete authoring environments, including their available models and tools.
Participants selected their general-purpose agents and model settings, while NarrativeSteward used a common model configuration.
Differences in representation, interaction support, familiarity, and story content also shaped the comparison.
The findings therefore concern the overall authoring experience; they do not isolate the contributions of model capability or individual system components.
Future studies could compare interface variants using the same model to examine the contribution of interface support more directly.

Second, the study involved \StudyCompleteN{} participants undertaking one authoring task in each condition.
The small sample and limited period of use constrain generalization to other author populations and longer creative projects.
Future studies could follow a broader range of authors over longer projects to examine how their review strategies, tool preferences, and revision practices change as they gain experience with the system.

Third, the evaluation emphasizes authors' perceived support and their accounts of making decisions, using study-specific items for the RQ-focused ratings.
Feature-helpfulness results for direct editing and whole-turn undo also reflect only a few respondents.
How these perceived benefits relate to the implementation of revision requests and the quality of the resulting stories requires further assessment.
Future studies could combine comparable behavioral observation in both environments with independent assessments of implemented revisions and completed narratives.

Finally, NarrativeSteward represents executable narratives using acyclic event and beat graphs, finite state-variable domains, and edge conditions involving only one variable each.
These constraints support complete reachable-state analysis within the modeled representation but restrict how recurring situations and complex conditions can be expressed.
The technical evaluation covered selected cases and scales, so verification costs for substantially larger or differently structured narratives remain to be examined.
Future work could extend the representation to support richer narrative forms and evaluate the computational cost of checking their execution.

\ifdefined\arxivversion\else
  \newpage
\fi
\section{Conclusion}
\label{sec:conclusion}

We presented NarrativeSteward, an interactive narrative authoring system that connects dialogue, structural review, and feedback through linked narrative artifacts to support authors in understanding and guiding autonomous agent implementation.
Technical tests validated the system's change records, recovery mechanisms, and execution diagnostics.
In a within-subject study with \StudyCompleteN{} participants, NarrativeSteward supported easier formulation of revision requests and inspection of changes, alongside greater perceived understanding of changes and story structure, than general-purpose agents.
Participants' experiences showed how structural review helped them develop subsequent requests and how they combined execution feedback with agent-led story analysis to direct further revisions.
These findings highlight the role of accessible, linked artifacts in sustaining author guidance across successive requests, and motivate authoring environments that accommodate different review needs and ways of creating while agents organize and carry out implementation.

\ifdefined\arxivversion
\else
\begin{acks}
Acknowledgments are omitted from the anonymized review PDF and will be
filled in the camera-ready version.
\end{acks}
\fi

\bibliographystyle{ACM-Reference-Format}
\bibliography{references}

\appendix

\section{Implementation Details}
\label{app:implementation}

NarrativeSteward uses a web client and Python backend to store project documents and typed JSON artifacts.
We implement the authoring agent using the Deep Agents framework~\cite{langchaindeepagents}, configured with narrative-specific instructions, tools, and a project-scoped file backend.
The backend exposes intent, outline, worldbuilding, event graphs, and per-event beat graphs for reading, searching, and editing, alongside read-only source materials.
These files supply both the agent’s working content and the artifacts presented in the authoring interface.

The main agent handles author dialogue and can inspect and edit project files directly.
It can also delegate substantial generation or rewriting to four specialized agents for outlines, worldbuilding, event graphs, and beat graphs.
Each receives role-specific instructions and can consult narrative schemas and guidance supplied as task-specific skills.
The agents share the project file backend, and the specialized agents inherit the main agent’s configured language model.
Subtask results return to the main agent, which can inspect the updated artifacts and determine how to continue the author’s request.
The study used the model configuration reported in Section~\ref{sec:procedure}, with no override of the provider’s temperature setting.

During a turn, the agents work on a shared isolated copy of the project; subtasks and file writes are serialized.
Structural validation checks artifact formats, references, and graph relationships, returning errors to the agent for correction before changes are saved together.
The main agent also has tools for retrieving the current execution-verification report and running verification when requested by the author.
These tools return structured diagnoses identifying affected narrative objects and conditions, which the agent can use when examining a problem and planning a revision.

Verification reports are associated with a content fingerprint computed from normalized event and beat graphs, including their text, world-card categories and identities, and markers for missing beat graphs.
Changes to these inputs invalidate the report.
Intent, outline, and world-card descriptions, names, and image paths are outside this fingerprint and do not invalidate it.

\section{User Study Materials and Supplementary Analyses}
\label{app:study-materials}

\subsection{Questionnaires}
\label{app:questionnaire}

This section presents the questionnaire items, response scales, and applicability conditions for the measures reported in this paper.
Condition-specific questions referred to the participant's work with NarrativeSteward or their selected general-purpose agent.

\subsubsection{Authoring Experience}

The thirteen measures in Table~\ref{tab:study_authoring_items} correspond to Table~\ref{tab:study_experience_summary} and are organized by research question.
Each uses a seven-point scale, with anchors shown for scores of 1, 4, and 7.
Depending on the item, participants could also select ``not experienced,'' ``cannot recall,'' or ``cannot judge.''

\begin{table*}[!t]
\caption{Questionnaire items for the thirteen authoring-experience measures compared in Table~\ref{tab:study_experience_summary}. Each question was asked for both tools. Anchors define scores of 1, 4, and 7 on seven-point scales.}\label{tab:study_authoring_items}
\centering\small
\setlength{\tabcolsep}{4pt}
\renewcommand{\arraystretch}{1.12}
\begin{tabular}{@{}>{\raggedright\arraybackslash}p{0.19\linewidth}>{\raggedright\arraybackslash}p{0.47\linewidth}>{\raggedright\arraybackslash}p{0.29\linewidth}@{}}
\toprule
Measure & Question & Anchors (1 / 4 / 7) \\
\midrule
\multicolumn{3}{@{}l}{\textbf{RQ1: Delegation experience and effort}}\\*
Formulating revisions & When considering further revisions to your work, how easy was it to identify what specifically you wanted to change? & Very difficult / Neither difficult nor easy / Very easy \\
Expressing requests & When asking the agent to revise the work, how easy was it to express your requirements clearly? & Very difficult / Neither difficult nor easy / Very easy \\
Inspecting changes & After the agent completed a revision, how easy was it to determine what it had actually changed? & Very difficult / Neither difficult nor easy / Very easy \\
Revision request fit & To what extent did the agent's result meet the revision requirements you had specified? & Not at all / Partially / Fully \\
\multicolumn{3}{@{}l}{\textbf{RQ2: Understanding and continued refinement}}\\*
Change understanding & To what extent did you understand the changes the agent made to the project? & Could not explain them at all / Could explain some of them / Could explain them fully \\
Edit stability & When you tried to change only one specific part, to what extent did unrelated content remain unchanged? & Not at all / Partially / Completely \\
Structure understanding & How well did you understand the work's overall narrative structure during the task? & Not at all / Partially / Very well \\
Locating ease & When you wanted to inspect a particular part of the content, how easy was it to find? & Very difficult / Neither difficult nor easy / Very easy \\
Refinement willingness & At the end of the task, how willing were you to continue refining this work? & Not at all willing / Moderately willing / Very strongly willing \\
\multicolumn{3}{@{}l}{\textbf{RQ3: Feedback-informed judgments and revision decisions}}\\*
Completion confidence & How confident were you that the current result met the task requirements? & Not at all confident / Somewhat confident / Very confident \\
Recovery ease & After a revision did not meet your expectations, how easy was it to restore or correct the work to a state from which you could continue creating? & Very difficult / Neither difficult nor easy / Very easy \\
Keep decision ease & After the agent completed a revision, how easy was it to decide whether to retain it? & Very difficult / Neither difficult nor easy / Very easy \\
Check decision ease & How easy was it to judge whether the current result needed further checking? & Very difficult / Neither difficult nor easy / Very easy \\
\bottomrule
\end{tabular}
\end{table*}

Formulating revisions applied to participants who had considered a change, even if they had not requested it.
Expressing requests required having made a request, inspecting changes required having examined the actual changes, and revision request fit required a result the participant could assess.

For each condition, participants also reported whether they had requested revisions, whether desired revisions remained unfinished, and what prevented them from completing those revisions.

\begin{table*}[!t]
\caption{Questions about revision experience and desired but unfinished revisions, asked for each tool. The reasons question allowed up to two selections from participants reporting unfinished revisions.}\label{tab:study_revision_items}
\centering\small
\setlength{\tabcolsep}{4pt}
\renewcommand{\arraystretch}{1.12}
\begin{tabular}{@{}>{\raggedright\arraybackslash}p{0.18\linewidth}>{\raggedright\arraybackslash}p{0.31\linewidth}>{\raggedright\arraybackslash}p{0.46\linewidth}@{}}
\toprule
Measure & Question & Response options \\
\midrule
Revision experience & After obtaining the first version of the work, which of the following best describes your experience? & Select one: did not consider further revisions; considered revisions but did not request them; requested revisions from the agent; could not recall; had not completed the original task; had not obtained a first version. \\
Unfinished revisions & At the end of the task, were there revisions you wanted to make but did not continue to complete? & Select one: yes; no; could not judge; could not recall. \\
Reasons for unfinished revisions & What factors led you not to complete these revisions? & Select up to two: the current result already met task needs; limited time or other activities; insufficient understanding of the existing content or structure to identify revisions; understanding the work but lacking a specific revision idea; anticipated effort to explain requirements or communicate repeatedly; anticipated effort to inspect revised results; concern about affecting satisfactory content; unwillingness to invest further at that time; another reason; inability to recall. \\
\bottomrule
\end{tabular}
\end{table*}

The reasons question applied only to those reporting unfinished revisions.
Selecting another reason allowed a written explanation; inability to recall was not combined with a substantive reason.

\subsubsection{Feature Use and Helpfulness}

For each NarrativeSteward feature in Table~\ref{tab:study_feature_items}, usage options were: used; aware but unused; unaware of the feature; or cannot recall.
Participants reporting use rated helpfulness for the stated purpose on a seven-point scale: 1 = not at all helpful, 4 = somewhat helpful, and 7 = extremely helpful.
They could also select cannot judge or cannot recall.

\begin{table*}[!t]
\caption{NarrativeSteward features and the purposes assessed by the helpfulness questions. Participants first reported feature use; those reporting use rated helpfulness from 1 (not at all helpful) to 7 (extremely helpful), with 4 indicating somewhat helpful.}\label{tab:study_feature_items}
\centering\small
\setlength{\tabcolsep}{4pt}
\renewcommand{\arraystretch}{1.12}
\begin{tabular}{@{}>{\raggedright\arraybackslash}p{0.3\linewidth}>{\raggedright\arraybackslash}p{0.66\linewidth}@{}}
\toprule
Feature & Purpose evaluated \\
\midrule
Structural view & Understanding the story structure through event and beat graphs. \\
Locate from changes & Finding the content modified by the agent. \\
Content inspection & Judging whether content matched the author's intent. \\
Direct editing & Making local revisions. \\
Whole-turn undo & Recovering from unsatisfactory agent revisions. \\
Execution verification report & Judging whether further checking or repair was needed. \\
Playtesting & Discovering content that needed revision. \\
\bottomrule
\end{tabular}
\end{table*}

\setcounter{topnumber}{1} 
\subsubsection{Tool Preferences and Usability}

\begin{table*}[!t]
\caption{Questions about overall tool preference and preferences for individual authoring activities. Participants selected one response for overall preference and one for each activity.}\label{tab:study_preference_items}
\centering\small
\setlength{\tabcolsep}{4pt}
\renewcommand{\arraystretch}{1.12}
\begin{tabular}{@{}>{\raggedright\arraybackslash}p{0.18\linewidth}>{\raggedright\arraybackslash}p{0.4\linewidth}>{\raggedright\arraybackslash}p{0.37\linewidth}@{}}
\toprule
Measure & Question & Response options \\
\midrule
Overall preference & If you undertook a similar interactive-narrative task again, which tool would you prefer to use? & Select one: definitely NarrativeSteward; probably NarrativeSteward; no clear preference; probably the selected general-purpose agent; definitely that agent; insufficient experience from the two tasks to judge. \\
Preference by authoring activity & Which approach would you choose for each of the following activities: developing a story direction; obtaining a first playable story; revising and polishing existing content; checking logic and preparing the final delivery? & Select one per activity: mainly NarrativeSteward; mainly the selected general-purpose agent; a combination of both; either tool; neither tool; insufficient experience to judge. \\
\bottomrule
\end{tabular}
\end{table*}

NarrativeSteward usability was measured using the published Simplified Chinese SUS wording \cite{wang2020chinesesus}, with ten items rated from 1 (strongly disagree) to 5 (strongly agree).

\subsubsection{Participant Background}

Background questions covered creative experience and roles, generative-AI use, narrative-editing-tool use, and experience with the selected general-purpose agent, including its model settings and available capabilities.

\subsection{Interview Guide}
\label{app:interview-guide}

Interview questions explored participants' experiences with both tools and the reasons for their judgments and choices.

\begin{enumerate}
  \item \textbf{Overall experience and creative direction.} Which parts of the two tasks felt most different? Which key creative decisions came from you and which from the agent? How did each work reflect the direction you wanted, and how did the tool, story topic, or task order affect your experience?
  \item \textbf{Delegating and expressing revisions.} Describe a particularly smooth or demanding implementation or revision, from making the request to deciding whether to accept the result. What did you want, how did you communicate it, and where did you invest effort? Include a request spanning several parts if applicable.
  \item \textbf{Finding and understanding changes.} Describe a time you noticed content needing revision. Where did you notice it, how did you find the relevant content, and how did you understand its relationship to the rest of the story? When revising one part, what happened to unrelated content?
  \item \textbf{Choosing how to revise or recover.} When did you ask the agent to revise, edit directly, undo, or leave the work unchanged? Describe how you reviewed the changes and recovered from an unsatisfactory result in each tool.
  \item \textbf{Checking the work and making decisions.} How did you judge whether the story could run as intended? Describe a decision to retain a result, request changes, continue checking, or stop checking. What information did you consult, what did you think it showed, and what did you do next? Did playtesting or verification change your revision plans? Was any information conflicting, unclear, or unhelpful?
  \item \textbf{Allocating decisions.} In a concrete example, which decisions should remain yours, which implementation work could be delegated to the agent, and which judgments could the system make using fixed rules?
  \item \textbf{Further work and tool preferences.} Were there changes you wanted but did not pursue, and why? Which tool would you choose for a longer project or at different authoring stages? What would you retain, remove, or combine from the two tools?
  \item \textbf{Additional experiences.} Is there anything else about your experience with either tool that we have not discussed?
\end{enumerate}

\subsection{Narrative Topics and Scale}
\label{app:story-scale}

The completed stories covered fantasy, historical fiction, science fiction, mystery, romance, and school-life settings.
Table~\ref{tab:study_story_scale} summarizes their scale using manually checked counts of main story stages, meaningful decision points, and reachable endings.
Stages correspond to major phases of the story, rather than individual graph nodes or pages.
Decision points are counted across all branches, including distinct locations on mutually exclusive routes; repeated visits to the same location count once.
Ending counts distinguish reachable narrative outcomes rather than the number of routes leading to them.

\begin{table*}[!t]
  \caption{Scale of the 24 completed interactive narratives, with 12 stories per condition.
NS denotes NarrativeSteward; Agent denotes participants' selected general-purpose agents.
Cells report mean [minimum, maximum] counts per story.
Stages are major narrative phases; decision points are distinct choice locations that affect content, later actions, or endings; endings are distinct reachable narrative outcomes.
Counts cover the whole story across its branches.}
  \label{tab:study_story_scale}
  \centering\small
\begin{tabular}{@{}lrrr@{}}
\toprule
Condition & Main story stages & Meaningful decision points & Reachable endings \\
\midrule
NS & 4.3 [4, 5] & 22.4 [8, 49] & 3.6 [3, 5] \\
Agent & 4.0 [3, 5] & 11.0 [5, 36] & 2.9 [2, 3] \\
\bottomrule
\end{tabular}

\end{table*}

\subsection{Supplementary Analyses and Results}
\label{app:paired_comparisons}

This section provides further details of the thirteen comparisons reported in Table~\ref{tab:study_experience_summary}.

Table~\ref{tab:study_comparison_statistics} supplements Table~\ref{tab:study_experience_summary} with medians, quartiles, and within-participant rating differences.
Both condition summaries use the same participants with valid ratings in both conditions for each measure.

\begin{table*}[!t]
  \caption{Paired rating distributions for the thirteen authoring-experience measures in Table~\ref{tab:study_experience_summary}. NS denotes NarrativeSteward; Agent denotes the selected general-purpose agent. Both condition summaries use the same eligible pairs (n). Mdn is the median, and Q1 and Q3 are the first and third quartiles. Paired difference is the median of the within-participant differences, calculated as NS minus Agent on the seven-point scale.}
  \label{tab:study_comparison_statistics}
  \centering\small
  \begingroup
\setlength{\tabcolsep}{3pt}
\renewcommand{\arraystretch}{1.1}
\begin{tabular}{@{}>{\raggedright\arraybackslash}p{0.055\linewidth}>{\raggedright\arraybackslash}p{0.25\linewidth}>{\raggedright\arraybackslash}p{0.075\linewidth}>{\raggedright\arraybackslash}p{0.19\linewidth}>{\raggedright\arraybackslash}p{0.19\linewidth}>{\raggedright\arraybackslash}p{0.165\linewidth}@{}}
\toprule
RQ & Measure & Paired $n$ & NS: Mdn [Q1, Q3] & Agent: Mdn [Q1, Q3] & Paired difference \\
\midrule
RQ1 & Formulating revisions & 12 & 6 [6, 7] & 3.5 [3, 4.25] & 2.5 \\
RQ1 & Expressing requests & 12 & 6 [5, 6.25] & 4 [4, 4.25] & 1.5 \\
RQ1 & Inspecting changes & 12 & 6.5 [6, 7] & 3 [2, 4] & 3.5 \\
RQ1 & Revision request fit & 12 & 5 [4, 5.25] & 4.5 [4, 5] & 0 \\
RQ2 & Change understanding & 12 & 6 [6, 7] & 4 [2.75, 5.25] & 1.5 \\
RQ2 & Edit stability & 8 & 6 [6, 7] & 4.5 [3.5, 5] & 2 \\
RQ2 & Structure understanding & 12 & 7 [6, 7] & 4 [4, 5] & 2.5 \\
RQ2 & Locating ease & 12 & 6 [6, 7] & 4 [3.75, 4] & 2 \\
RQ2 & Refinement willingness & 12 & 5 [4, 6] & 4 [3.5, 5] & 1 \\
RQ3 & Completion confidence & 12 & 6 [5, 7] & 5 [4.75, 5] & 1.5 \\
RQ3 & Recovery ease & 8 & 6 [5.75, 7] & 4.5 [3.75, 5.25] & 2 \\
RQ3 & Keep decision ease & 10 & 6 [5, 6] & 3.5 [2.25, 4] & 2 \\
RQ3 & Check decision ease & 9 & 5 [5, 6] & 3 [2, 4] & 2 \\
\bottomrule
\end{tabular}
\endgroup

\end{table*}

Quartiles were calculated using linear interpolation.
The Wilcoxon tests in Table~\ref{tab:study_experience_summary} were computed from within-participant rating differences.
Zero differences were omitted from the test calculation, and equal absolute differences received average ranks.
The p-values were obtained by enumerating the possible positive and negative sign assignments to the nonzero differences.
The tests assume independent participants and, under the null hypothesis, differences symmetric about zero.

\FloatBarrier
\section{Execution Verification and Technical Evaluation Details}
\label{app:technical-evaluation}

This appendix describes the execution verification algorithm introduced in Section~\ref{sec:verification}, followed by the technical test cases and measurements summarized in Section~\ref{sec:tech-eval}.
Tests used predefined inputs in isolated project copies and exercised the implemented change, recovery, and verification mechanisms without invoking a language model.

\subsection{Execution Verification Algorithm}
\label{sec:verification-algorithm}

\textbf{Execution positions.} The system expands the event graph and its beat graphs into one directed acyclic execution graph.
Each event has an entry position, positions for its beats, and a position for selecting the next event after the beat graph finishes.
Entering an event leads to its entry beat; a beat with no outgoing beat edge leads to the event-selection position.
Beat effects execute on entry, before outgoing conditions are evaluated; event-edge conditions are evaluated after the beat graph finishes.
Completing an ending event terminates execution.
This representation lets the analysis follow variable values across both graph levels, including values set in an early beat and tested in a later event.

\textbf{Variables needed for later conditions.} For each position \(p\), let \(L_{\mathrm{before}}(p)\) and \(L_{\mathrm{after}}(p)\) denote the variables whose values must be retained before and after its effects.
The system computes these sets in reverse topological order, visiting successors before their predecessors.
For each outgoing transition \(p\to r\), it includes the variable tested by that transition, if any, and the variables needed before the successor’s effects:

\[
L_{\mathrm{after}}(p)
= \bigcup_{p\to r}\left(\operatorname{read}(p\to r)\cup L_{\mathrm{before}}(r)\right).
\]

Here, \(\operatorname{read}\) is empty for an unconditional transition and otherwise contains its condition variable; the union is empty at a position with no outgoing transitions.
To obtain \(L_{\mathrm{before}}(p)\), the system starts with \(L_{\mathrm{after}}(p)\) and processes the position’s effects in reverse order.
A fixed assignment \texttt{set x = c} removes \texttt{x} from the required set because it overwrites the incoming value.
An update \texttt{add x by c} leaves the set unchanged: when the resulting \texttt{x} is needed, its incoming value is also needed.
These operations use constants and do not read other variables.
Taking the union over all outgoing transitions retains a variable needed on any branch, even before the analysis determines which branches are reachable.

\textbf{State propagation and diagnostics.} The system starts at the entry event with the declared initial values and processes positions in topological order.
At each position, it stores distinct joint assignments to \(L_{\mathrm{before}}(p)\).
For each assignment, it applies effects in their declared order, including the bounds on integer updates, retains \(L_{\mathrm{after}}(p)\), and tests every outgoing condition.
Each enabled transition records edge coverage and forwards the values needed by its successor; an identical assignment already present there is not added again.
Joint assignments are kept intact, preserving correlations among variables reached along the same execution.
The retained values are packed into an integer using separate bit fields, and each position’s state set is released after processing.
The system records reached events and beats, enabled edges, and reachable non-ending states with no continuation.
For reported dead-end states, a targeted traversal reconstructs an execution from the entry to the blocked state, providing the choices and conditions used to explain the diagnosis.

\textbf{Why merging preserves the checks.} If two states agree on \(L_{\mathrm{before}}(p)\), their effects produce equal values for every variable in \(L_{\mathrm{after}}(p)\).
Fixed assignments produce the same values, and relevant additions start from equal values and use the same increments and bounds.
Their outgoing conditions therefore have the same outcomes, and each enabled transition produces equal retained values at its successor.
Repeating this argument along the acyclic graph shows that merged states have the same possible future transitions and termination behavior.
Recording visited objects and enabled edges before eliminating duplicate successor assignments preserves coverage from different incoming routes.
Consequently, merging preserves event and beat reachability, whether each edge is ever enabled, and the presence of reachable dead ends under the represented execution rules.

\subsection{Change Records and Recovery}
\label{sec:tech-eval-procedure}

The four editing cases covered an event-condition edit; a batch changing beat text, an effect, and an edge; the addition of a world card, an event, and its beat graph; and a text edit with asset addition and removal.
Change detection compares identified details with expected details, precision measures the proportion of reported details that were expected, and localization checks whether each locatable detail points to the correct narrative object.

Recovery cases introduced an exception, a timeout, a recursion-limit exception, or a user stop after writes to project content and verification state.
Each tested whether turn finalization restored the pre-turn project without producing a successful changeset.
Undo cases covered an event field, a batch within a beat graph, and cross-layer additions including an asset.
Checks covered restoration of prior content, removal of added assets, and unchanged content on repeated undo.
Conflict cases introduced a later content edit, a revision-only change, or a conflicting fragment within a cross-layer changeset, testing rejection of the entire undo while retaining subsequent work.

\subsection{Execution Verification Tests}
\label{sec:tech-eval-tq2-correctness}

Three cases with hand-computed outcomes tested linear flow, clamping values to their declared bounds, and overwriting values after branches converge.
We prepared three pairs of faulty and corrected projects, covering an unreachable event, a never-enabled edge whose target was reachable through another route, and a reachable dead-end state.
Each pair had an expected issue type and location for the faulty version and a corrected version expected to pass execution verification.
Report-applicability tests covered five edits to checked inputs, including conditions, effects, beat-graph edges, and speaker-card identities, and two edits to world-card descriptions and image paths.

\subsection{Runtime and Resource Measurements}
\label{sec:tech-eval-tq2-resources}

The resource cases comprised an integrated reference story and four generated structures targeting different sources of verification cost.
The reference story contained seven events, 34 beats, and five state variables.
The generated cases used 12 binary-choice stages with a scalar range of 0--4095, ten Boolean variables relevant to later conditions, 32 branches, and a 256-beat chain with a fault near its end, respectively.
Measurements ran on Linux with an AMD EPYC 7K83 processor and Python 3.12.4.
Each case received one warmup and five measured runs, each in a fresh process.
Verification time included reading the saved project, structural checks, state propagation, any required diagnostic-route reconstruction, and report generation.
Peak memory was measured as the process's peak resident set size using Linux \texttt{VmHWM}.
Table~\ref{tab:tech_eval_tq2_resources} reports verification time, peak memory, and the number of explored execution states.
All runs completed; the distant-fault case required reconstruction of a route leading to a never-enabled edge.

\begin{table*}[!t]
  \caption{Execution verification resources for one integrated reference story and four generated structures.
Each case received one warmup and five measured runs in separate processes.
Time is reported as median [minimum, maximum]; peak memory and explored execution states are medians.
The distant-fault case completed with the expected diagnosis of a never-enabled edge.}
  \label{tab:tech_eval_tq2_resources}
  \centering\small
\begin{tabular}{lrrr}
\toprule
Case & Verification time (ms) & Peak memory (MiB) & Execution states \\
\midrule
Integrated reference & 3.5 [3.3, 3.6] & 36.8 & 136 \\
Large scalar domain & 59.4 [57.0, 60.5] & 37.5 & 45,051 \\
Many relevant Boolean variables & 20.7 [20.2, 22.1] & 37.3 & 16,374 \\
Wide branch & 5.9 [5.6, 6.7] & 37.0 & 199 \\
Distant fault & 10.4 [10.1, 11.3] & 37.9 & 261 \\
\bottomrule
\end{tabular}

\end{table*}

\end{document}